\documentclass{SciPost}

\hypersetup{
    colorlinks,
    linkcolor={red!50!black},
    citecolor={blue!50!black},
    urlcolor={blue!80!black}
}

\usepackage[bitstream-charter]{mathdesign}
\DeclareSymbolFont{usualmathcal}{OMS}{cmsy}{m}{n}
\DeclareSymbolFontAlphabet{\mathcal}{usualmathcal}

\fancypagestyle{SPstyle}{
\fancyhf{}
\lhead{\colorbox{scipostblue}{\bf \color{white} ~SciPost Physics }}
\rhead{{\bf \color{scipostdeepblue} ~Submission }}

\fancyfoot[C]{\textbf{\thepage}}
}

\usepackage{dcolumn}
\usepackage{bm}
\usepackage{multirow}
\usepackage{booktabs}
\newcommand{\ra}[1]{\renewcommand{\arraystretch}{#1}}

\usepackage{physics}
\usepackage{dsfont}
\usepackage{tikz}
\usetikzlibrary{decorations.pathmorphing}
\usetikzlibrary{decorations.pathreplacing}
\definecolor{green}{rgb}{0, 0.55, 0.45}
\definecolor{orange}{rgb}{1, 0.66, 0}
\definecolor{anthracite}{gray}{0.6}

\newcommand{\hop}{H}
\renewcommand{\op}[2]{#1_{#2}}
\newcommand{\hc}[2]{#1_{#2}^{\dagger}}

\begin{document}

\pagestyle{SPstyle}

\begin{center}{\Large \textbf{\color{scipostdeepblue}{
The Ising dual-reflection interface: \\  $\mathbb{Z}_4$ symmetry and Majorana strong zero modes\\
}}}\end{center}

\begin{center}\textbf{
Juliane Graf\textsuperscript{1},
Federica M. Surace\textsuperscript{2},
Marcus Berg\textsuperscript{1,3} and
Sergej Moroz\textsuperscript{1,3}
}\end{center}

\begin{center}
{\bf 1} Department of Engineering and Physics, Karlstad University, Karlstad, Sweden
\\
{\bf 2} Department of Physics and Institute for Quantum Information and Matter, California Institute of Technology, Pasadena, California 91125, USA
\\
{\bf 3} Nordita, KTH Royal Institute of Technology and Stockholm University, Stockholm, Sweden
\end{center}

\section*{\color{scipostdeepblue}{Abstract}}
\textbf{\boldmath{%
We investigate an interface in the transverse field quantum Ising chain connecting an ordered ferromagnetic phase and a disordered paramagnetic phase that are Kramers-Wannier duals of each other. Unlike prior studies focused on non-invertible defects, this interface exhibits a symmetry that combines Kramers-Wannier transformation with spatial reflection. We demonstrate that, under open boundary conditions, this setup gives rise to a discrete $\mathbb{Z}_4$ symmetry, encompassing the conventional $\mathbb{Z}_2$ Ising parity as a subgroup, while in a closed geometry a non-invertible symmetry emerges. Using the Jordan-Wigner transformation, we map the spin chain onto a solvable quadratic Majorana fermion system. In this formulation, the $\mathbb{Z}_4$ symmetry is realized manifestly as a parity-dependent reflection with respect to a Majorana site, in contrast to the conventional reflection which mirrors with respect to the central link of the Majorana chain.  Additionally, we construct Majorana strong zero modes that retain the $\mathbb{Z}_4$ symmetry, ensure degeneracies of all energy eigenstates, and are robust under generic local symmetry-preserving perturbations  of the fermion model, including interactions. Finally, we develop quantum circuit realizations of our model paving the way towards the creation of exact Majorana strong zero modes with digital quantum hardware.
}}

\vspace{\baselineskip}

\noindent\textcolor{white!90!black}{%
\fbox{\parbox{0.975\linewidth}{%
\textcolor{white!40!black}{\begin{tabular}{lr}%
  \begin{minipage}{0.6\textwidth}%
    {\small Copyright attribution to authors. \newline
    This work is a submission to SciPost Physics. \newline
    License information to appear upon publication. \newline
    Publication information to appear upon publication.}
  \end{minipage} & \begin{minipage}{0.4\textwidth}
    {\small Received Date \newline Accepted Date \newline Published Date}%
  \end{minipage}
\end{tabular}}
}}
}


\vspace{10pt}
\noindent\rule{\textwidth}{1pt}
\tableofcontents
\noindent\rule{\textwidth}{1pt}
\vspace{10pt}


\section{Introduction}
\label{sec:intro}
Over the past decade, our understanding of symmetries has evolved significantly, providing deep new insights into what constitutes a symmetry and how it constrains physical observables. This modern approach has unified various aspects of complex physical systems, leading to the new concept of generalized symmetries, for reviews see for example \cite{nandkishore2019fractons, pretko2020fracton, mcgreevy2023generalized,  shao2023s, gomes2023introduction, bhardwaj2024lectures, luo2024lecture, gromov2024colloquium, schafer2024ictp}. Alongside symmetries, dualities play an important role for our theoretical understanding of a wide range of physics. Dualities relate two distinct theories that nonetheless yield identical physical predictions.
A prototypical example is the duality of the transverse field Ising chain - the famous Kramers-Wannier duality which relates a ferromagnet to a paramagnet. While often complementary to each other, the notions of symmetry and duality can become intertwined within the recent more general understanding of symmetry. In particular, it has been shown that the Kramers-Wannier duality transformation can be implemented by a non-invertible operator that commutes with the Ising Hamiltonian at criticality.
 
A useful modern perspective is to view symmetries as the operators associated with  topological defects 
\cite{affleck1997boundary, PhysRevLett.93.070601, grimm2002spectrum, frohlich2007duality, aasen2016topological, aasen2020topological, PhysRevB.94.115125, seiberg2024majorana, seiberg2024non, carqueville2023topological}.
The location of such defects does not change the properties of the model since they can be displaced along the system by a local unitary transformation. In particular, Refs. \cite{aasen2016topological, aasen2020topological, seiberg2024majorana, seiberg2024non} investigated a non-invertible Kramers-Wannier symmetry and explicitly constructed corresponding topological defects and their fusion rules in the two-dimensional classical Ising model and the equivalent quantum one-dimensional spin chain. Kramers-Wannier non-invertible defects arise as interfaces between dual regions \cite{PhysRevLett.93.070601, grimm2002spectrum, frohlich2007duality, aasen2016topological, aasen2020topological, PhysRevB.94.115125, seiberg2024majorana, seiberg2024non}, for a recent review see \cite{carqueville2023topological}.
It was found in \cite{aasen2016topological} that such an interface in the quantum Ising chain, respectively the Kitaev chain, can host an exact strong zero mode, a zero energy excitation causing a degeneracy of all energy states (as opposed to  ``weak'' zero modes causing ground state degeneracies alone). This strong zero mode is exact in the sense that its commutation with the Hamiltonian does not require the thermodynamic limit but persists for chains of an arbitrary, finite length.

Strong zero modes can exist at both system boundaries, where they were first discovered as edge modes of the Kitaev chain \cite{kitaev2001unpaired}, and phase boundaries \cite{aasen2016topological,olund2023boundary, yan_duality_2025}. In the Kitaev chain, strong zero modes are guaranteed to exist at the interface between a trivial and a topological phase, while the scenario is more complicated for the corresponding Ising model
\cite{olund2023boundary}. Strong zero modes can retain memory of the initial state over parametrically long times \cite{kemp2017long}, even under perturbations, which renders them well suited for experimental realizations in noisy-intermediate scale quantum simulators \cite{Google2022}.

\begin{figure}[t]
    \centering
    
    	\begin{tikzpicture}[scale=0.9]
		\tikzstyle{every node}=[font=\fontsize{14}{14}\selectfont]
		
		\foreach \x in {-6, -4, -2}{
			\draw[green] (\x, -0.5) node[anchor=north]{$h$};
			\draw[ green, line width=3pt] (\x,0.5) -- (\x,-0.5);
		}
		
		\draw[anthracite] (0, -0.5) node[anchor=north]{$J_0$};
		\draw[line width=3pt, anthracite] (0,0.5) -- (0,-0.5);
		
		\foreach \x in {2, 4}{
			\draw[orange] (\x, -0.5) node[anchor=north]{$J$};
			\draw[ orange, line width=3pt] (\x,0.5) -- (\x,-0.5);
			
		}
		
		\foreach \x in {-5, -3}{
			\draw[orange] (\x, 0.25) node[anchor=south]{$J$};
			\draw[decorate, decoration={snake, segment length=12}, orange, line width=3pt] (\x-1,0) -- (\x+1,0);
			
		}
		\draw[anthracite] (-1, 0.25) node[anchor=south]{$J_0$};
		\draw[decorate, decoration={snake, segment length=12}, line width=3pt, anthracite] (-2,0) -- (0,0);
		\foreach \x in {3, 1}{
			\draw[green] (\x, 0.25) node[anchor=south]{$h$};
			\draw[decorate, decoration={snake, , segment length=12}, green, line width=3pt] (\x-1,0) -- (\x+1,0);
		}
		
		\draw (-6.5,0) node{...};
		\draw (4.5,0) node{...};
		\foreach \x in {-6,-4,-2,...,4}
		{        
			\filldraw (\x,0) circle (0.1);
		}
		\begin{scope}[shift={(6.5,0.5)}, scale=0.6]
			\draw[decorate, decoration={snake, segment length = 11.5}, black, line width=2pt] (-1, 0) -- (1,0);
			\draw (1.5,0) node[anchor=west]{$X_j X_{j+1}$};
	
		\end{scope}
		\begin{scope}[shift={(6.5,-0.25)}, scale=0.6]
			\draw[ black, line width=2pt] (0, 0.5) -- (0,-0.5);
			\draw (1.5,0) node[anchor=west]{$Z_j$};

		\end{scope}
        \draw[] (5.7,-0.75) rectangle (9.7,1.1);

	\end{tikzpicture}
 \caption{An interface between ferromagnetic and paramagnetic Ising regions that are Kramers-Wannier duals of each other: The strengths of the Ising couplings (horizontal wavy lines) and transverse field couplings (vertical straight lines)  are color-coded. By virtue of the duality, the Ising and transverse field couplings on opposite sides of the interface have the same strength. The interface Hamiltonian is a sum of the central link-Ising coupling and the  transverse field acting on the spin to the right of the interface, $ H_{\text{int}} = -J_0\, (\op{X}{n}\op{X}{n+1} + \op{Z}{n+1}) \,.$ }
    \label{fig:KWi}
\end{figure}

In the focus of this paper are symmetries and strong zero modes of interfaces between an Ising ferromagnet and paramagnet.
Specifically, we design and investigate an interface between Kramers-Wannier duals of the transverse field Ising model that is illustrated in Fig.~\ref{fig:KWi}. The interface is not topological but inspired by symmetry arguments instead. Specifically, we demand that a composition\footnote{The same composition of the Kramers-Wannier transformation and spatial reflection was recently identified as a non-invertible reflection symmetry of translation-invariant lattice models with modulated symmetries \cite{pace_gauging_2025}. It was also introduced as an alternative implementation of the Kramers-Wannier duality that does not mix with lattice translation \cite{gorantla_tensor_2025}.} of the Kramers-Wannier transformation and spatial reflection about the interface constitute a symmetry, as illustrated schematically for an open chain in Fig.~\ref{fig:KW-r}. This symmetry is valid even away from the Ising critical point, unlike the Kramers-Wannier symmetry.

The first part of this paper, Secs.~\ref{sec:spin} and ~\ref{sec:ferm}, studies the interpretation of our symmetry in the spin- and fermion language as well as its symmetry algebra. In Sec.~\ref{sec:spin} we discover that, for open boundary conditions, this operation generates a discrete $\mathbb{Z}_4$ symmetry\footnote{More precisely, the operation constitutes a self-duality of order four, also referred to as self-quadrality. However, for simplicity, we will refer to it as symmetry in the following.} which contains the conventional Ising parity $\mathbb{Z}_2$ symmetry as a subgroup. When we use closed boundary conditions, the $\mathbb{Z}_4$ symmetry is lost, giving way to a new, non-invertible symmetry that, in contrast to the Kramers-Wannier non-invertible symmetry, is valid even away from criticality. In Sec.~\ref{sec:ferm} we apply the Jordan-Wigner transformation to the spin chain, which gives rise to a solvable quadratic Majorana problem. In this language, we discover in an open geometry that the identified $\mathbb{Z}_4$ symmetry generator acts on individual Majorana fermions as parity-dependent reflection with respect to a next-to-middle Majorana site, see Fig.~\ref{fig:Majorana_chain}. Such a dual reflection should be contrasted with an ordinary reflection transformation, which, in the fermion formulation, mirrors with respect to the central link of the Majorana chain. In the light of the above, we refer to the studied system as an Ising dual-reflection interface. We also discuss how a $\mathbb{Z}_4$ symmetric fermion model can be constructed in a closed geometry.

In the remainder of the paper we reveal how dual-reflection symmetry entails Majorana strong zero modes \cite{alicea2016topological, fendley2016strong} (Sec.~\ref{sec:SMZM}) and Floquet Majorana strong zero modes (Sec.~\ref{sec:quantum circuit}), i.e., localized operators that commute (up to corrections exponentially small in system size) with the Hamiltonian or time evolution operator, but toggle the fermion parity. Such operators ensure degeneracies not only of the low-energy eigenstates, but across the whole spectrum.

Dual-reflection symmetric chains host edge modes as well as interface modes, whose operators and $\mathbb{Z}_4$ symmetry transformation we construct explicitly. Among them are exact Majorana strong zero modes which cannot be lifted by any local interaction term that preserves the $\mathbb{Z}_4$ symmetry. These special modes are symmetry-protected excitations that remain pinned to exactly  zero energy in finite open chains, in contrast to usual zero modes which are split in energy away from the thermodynamic limit. Our construction allows us to identify two regimes of the studied model which have distinct degeneracies in the open geometry. In Sec.~\ref{sec:quantum circuit} we extend our constructions to the framework of discrete-time Floquet evolution of quantum circuits. We propose $\mathbb{Z}_4$ symmetric unitary quantum circuits in the spin and fermion language and construct exact Floquet Majorana strong zero modes protected by the symmetry. Finally, we summarize  several promising future directions in Sec.~\ref{sec:outlook}.

The appendices contain background information and alternative derivations of selected results outlined in the main text. In Appendix~\ref{appA} we introduce the Kramers-Wannier transformation in the spin language, and in Appendix~\ref{appB} we derive the expressions for the Kramers-Wannier and spatial reflection transformations in the fermion formulation. We discuss an alternative definition of the Kramers-Wannier transformation and how it alters the dual-reflection interface in Appendix~\ref{app:alternative_KW}. In Appendix~\ref{app:BdG} we present the Bogoliubov-de Gennes equations, which can be used to solve a generic quadratic fermion Hamiltonian, applied to the dual-reflection interface. In Appendix~\ref{app:SZMs} we construct the strong zero modes by solving the Bogoliubov-de Gennes equations. Appendix~\ref{app:FMZM} provides details on the derivation of the Floquet strong zero modes.

\section{The Ising dual-reflection interface and its symmetry} \label{sec:spin}
This section contains a detailed construction of the dual-reflection symmetric Ising spin chain in both open and closed geometry. We then formulate the dual-reflection symmetry operator and study its symmetry algebra.

\subsection{Setup}

In this paper we investigate an interface in the transverse field Ising model, illustrated in Fig.~\ref{fig:KWi}, between the ordered ferromagnetic and disordered paramagnetic phase. The left and right parts are fine-tuned to be Kramers-Wannier duals\footnote{The Kramers-Wannier duality relates two Ising chains with interchanged Ising nearest-neighbour and transverse field coupling. A mathematically precise formulation is discussed below and in Appendix~\ref{appA}. } of each other. The interface part of the Hamiltonian contains the defect Ising coupling $J_0$ between the two interface spins, and in addition a transverse field term of the same strength $J_0$ acting on only one of the spins.\footnote{Which spin it is, left or right, depends on the convention chosen for the Kramers-Wannier transformation, and on whether the chain is comprised of an even or odd number of sites. For our choice, the defect transverse field $J_0$ acts on the right spin, see Fig.~\ref{fig:KWi}. In Appendix~\ref{app:alternative_KW} we discuss an alternative implementation which, for an even number of sites, requires that the defect Hamiltonian contains the field acting on the left spin.} Consequently, it is not reflection symmetric. Our interface Hamiltonian also differs from the topological non-invertible defect discussed extensively in the lattice Ising model \cite{aasen2016topological,seiberg2024majorana, seiberg2024non}. As we will demonstrate in the following, the guiding principle for our model is the emergence of a new symmetry which is composed of the Kramers-Wannier transformation and spatial reflection, as schematically illustrated in Fig.~\ref{fig:KW-r}.

For concreteness, but without loss of generality, we assume that the spin chain consists of an even number of sites $N=2n$. On a finite open chain, the spin Hamiltonian illustrated in Fig.~\ref{fig:KWi} can be written as
\begin{align} \label{H_int}
	&\hop = - \sum_{j=1}^{N-1} J_j\, \op{X}{j}\op{X}{j+1} -  \sum_{j=2}^N h_j \,\op{Z}{j}\,,\\
	&J_j = \begin{cases}
		J & j = 1, \,\dots,\, n-1 \\
		J_0 & j = n\\
		h & j = n+1,\,\dots,\,N-1
	\end{cases}\,,  \nonumber  \\
	&h_j = \begin{cases}
		h & j = 2, \,\dots,\, n \\
		J_0 & j = n+1\\
		J & j = n+2,\,\dots,\,N
	\end{cases}\,, \nonumber
\end{align}
where $\op{X}{j}$ and $\op{Z}{j}$ are Pauli matrices acting on sites labeled by $j=1,2,\dots, N$. The interface coupling $J_0$ can be chosen arbitrarily. In the special case $J=h=J_0\,,$ the critical Ising model is recovered. We set $h_{j=1} = 0$ as will be explained below.

\begin{figure}[t]
    \centering

    \begin{tikzpicture}[scale=0.7]
	\tikzstyle{every node}=[font=\fontsize{14}{14}\selectfont]
	
	\foreach \x in {-6, -4, -2}{
		\draw[green] (\x, -0.5) node[anchor=north]{$h$};
		\draw[ green, line width=3pt] (\x,0.5) -- (\x,-0.5);
	}
	
	\draw[anthracite] (0, -0.5) node[anchor=north]{$J_0$};
	\draw[line width=3pt, anthracite] (0,0.5) -- (0,-0.5);
	
	\foreach \x in {2, 4}{
		\draw[orange] (\x, -0.5) node[anchor=north]{$J$};
		\draw[ orange, line width=3pt] (\x,0.5) -- (\x,-0.5);
		
	}
	
	\foreach \x in {-5, -3}{
		\draw[orange] (\x, 0.25) node[anchor=south]{$J$};
		\draw[decorate, decoration={snake, segment length=12}, orange, line width=3pt] (\x-1,0) -- (\x+1,0);
		
	}
	\draw[anthracite] (-1, 0.25) node[anchor=south]{$J_0$};
	\draw[decorate, decoration={snake, segment length=12}, line width=3pt, anthracite] (-2,0) -- (0,0);
	\foreach \x in {3, 1}{
		\draw[green] (\x, 0.25) node[anchor=south]{$h$};
		\draw[decorate, decoration={snake, , segment length=12}, green, line width=3pt] (\x-1,0) -- (\x+1,0);
	}
	
	\draw (-6.5,0) node{...};
	\draw (4.5,0) node{...};
	\foreach \x in {-6,-4,...,4}
	{        
		\filldraw (\x,0) circle (0.1);
	}
	\begin{scope}[shift={(0,-3)}]
		\foreach \x in {-6, -4}{
			\draw[orange, line width=3pt] (\x,0.5) -- (\x,-0.5);
		}
		
		\draw[anthracite, line width=3pt] (-2,0.5) -- (-2,-0.5);
		
		\foreach \x in {0, 2, 4}{
			\draw[green, line width=3pt] (\x,0.5) -- (\x,-0.5);
			
		}
		
		\foreach \x in {-5,-3}{
			\draw[decorate, decoration={snake, , segment length=12}, green, line width=3pt] (\x-1,0) -- (\x+1,0);
			
		}
		\foreach \x in {3, 1}{
			\draw[decorate, decoration={snake, , segment length=12}, orange, line width=3pt] (\x-1,0) -- (\x+1,0);
		}
		
		\draw (-6.5,0) node{...};
		\draw (4.5,0) node{...};
		\draw[decorate, decoration={snake, segment length=12}, line width=3pt, anthracite] (-2,0) -- (0,0);
		\foreach \x in {-6,-4,-2,...,4}
		{        
			\filldraw (\x,0) circle (0.1);
		}
	\end{scope}
	\begin{scope}[shift={(0,-6)}]
		\foreach \x in {-6, -4, -2}{
			\draw[green, line width=3pt] (\x,0.5) -- (\x,-0.5);
		}
		
		\draw[anthracite, line width=3pt] (0,0.5) -- (0,-0.5);
		
		\foreach \x in {2, 4}{
			\draw[orange, line width=3pt] (\x,0.5) -- (\x,-0.5);
		}
		
		\foreach \x in {-5, -3}{
			\draw[decorate, decoration={snake, , segment length=12}, orange, line width=3pt] (\x-1,0) -- (\x+1,0);
		}
		
		\foreach \x in {3, 1}{
			\draw[decorate, decoration={snake, , segment length=12}, green, line width=3pt] (\x-1,0) -- (\x+1,0);
		}
		
		\draw (-6.5,0) node{...};
		\draw (4.5,0) node{...};
		\draw[decorate, decoration={snake, segment length=12}, line width=3pt, anthracite] (-2,0) -- (0,0);
		\foreach \x in {-6,-4,-2,...,4}
		{        
			\filldraw (\x,0) circle (0.1);
		}
	\end{scope}
	\begin{scope}[shift={(-1,0)}]
		\draw[->, very thick] (-7,0)..controls (-8,-0.5)  and (-8,-2.5)..(-7, -2.9);
		\draw[->, very thick] (-7,-3.1)..controls (-8,-3.5)  and (-8,-5.5)..(-7, -6);
		\draw (-9, -1.5) node[anchor=east, text width=2cm]{Kramers-Wannier};
		\draw (-9, -4.5) node[anchor=east]{reflection};
		\draw[anthracite, line width=3pt, decorate, decoration={snake, segment length=12.5}] (-11.5, -5.5)--(-9.5, -5.5);
		\draw[blue, dashed, line width=2pt] (-10.5, -4.9)--(-10.5,-6.1);
		
	\end{scope}
\end{tikzpicture}
    \caption{Schematic action of the dual-reflection symmetry $S$
    on the Hamiltonian \eqref{H_int} of the dual-reflection interface: In the first step, the Kramers-Wannier transformation locally interchanges the nearest-neighbor Ising coupling with the external field strength. The second step is a spatial reflection. Given an even number of sites, we reflect about the central link.}
    \label{fig:KW-r}
\end{figure}

\subsection{ $\mathbb{Z}_4$ symmetry in the open chain}
It is clear that the model enjoys the ordinary Ising $\mathbb{Z}_2$ symmetry generated by $Q=\prod_{j=1}^{N} \op{Z}{j}$. We will now show that there exists a more elementary symmetry in our model, for arbitrary values of the couplings $h$, $J$ and $J_0\,.$ To reveal this, we first consider the unitary operator $\op{U}{KW}$, see \cite{shao2023s, seiberg2024non} and Appendices~\ref{appA} and~\ref{app:alternative_KW},\footnote{The index $j$ runs from $N$ to $2$ from left to right.}
\begin{equation} \label{KW}
    \op{U}{KW}=\,e^{-2\pi i N/8}  \left(\prod_{j=N}^{2} \frac{1+i \op{Z}{j}}{\sqrt{2}} \frac{1+i \op{X}{j} \op{X}{j-1}}{\sqrt{2}} \right) \frac{1+i \op{Z}{1}}{\sqrt{2}}\,,
\end{equation}
which implements the Kramers-Wannier transformation as follows:
\begin{equation}
\begin{split}\label{KW1}
    &{U}_{KW} \, \op{Z}{1} \, {U}_{KW}^{-1} = Q \op{X}{1} \op{X}{N}, \quad \\
    &{U}_{KW} \, \op{X}{1} \op{X}{N} \,{U}_{KW}^{-1} =  Q \op{Z}{N},
\end{split}
\end{equation}
and
\begin{equation} \label{KW2}
\begin{split}
   &{U}_{KW} \,  \op{Z}{j} \, {U}_{KW}^{-1} = \op{X}{j-1} \op{X}{j}\,, \quad \\
   &U_{KW}\, \op{X}{j-1} \op{X}{j} \, U_{KW}^{-1} = \op{Z}{j-1}
\end{split}
\end{equation}
for $j=2,3,\dots, N$.
In the transverse field Ising model with constant parameters, this operation maps between ferromagnetic and paramagnetic regimes called dual to each other.  Under the Kramers-Wannier duality, the critical model ($J=h$) is invariant.

We concatenate this transformation with the spatial reflection $R$ that reflects with respect to the central link and thus swaps the sites $j\leftrightarrow N+1-j$. In the Hilbert space of the spin chain, the reflection is implemented as a product of $n$ two-qubit SWAP gates.
The combined operation of Kramers-Wannier transformation and spatial reflection, $S=R \op{U}{KW}$, commutes with the Hamiltonian \eqref{H_int}, as illustrated in Fig.~\ref{fig:KW-r}. We therefore refer to this model as (Ising) dual-reflection interface. As the transformation $S=R \op{U}{KW}$ maps $Z_1$ on the non-local operator $Q X_N X_1$, in the open chain, the local $S$-symmetric Hamiltonian must have a vanishing transverse field on the first spin ($h_{j=1}=0$).

An investigation of the symmetry $S$ and its consequences is the main purpose of this study. Since the reflection $R$ and the Kramers-Wannier transformation $\op{U}{KW}$ both independently commute with the Ising symmetry $Q$, $S$ also commutes with $Q$.
We will now prove that $S$ generates the Ising symmetry $Q$, namely $S^2=Q$.
Taking advantage of the fact that $R^{-1}=R$ we find
\begin{align}
    S^2=&\,RU_{KW}RU_{KW}=(RU_{KW}R^{-1})U_{KW} \nonumber\\
    =&\,e^{-4\pi i N/8} \prod_{j=1}^{N-1}  \pqty{ \frac{1+i \op{Z}{j}}{\sqrt{2}} \frac{1+i \op{X}{j} \op{X}{j+1}}{\sqrt{2}} } \frac{1+i \op{Z}{N}}{\sqrt{2}} \prod_{j=N}^{2}  \pqty{ \frac{1+i \op{Z}{j}}{\sqrt{2}} \frac{1+i \op{X}{j} \op{X}{j-1}}{\sqrt{2}} } \frac{1+i \op{Z}{1}}{\sqrt{2}}\,.\label{eq:product}
\end{align}
We can evaluate this product using that
\begin{equation}
    \left(\frac{1+iZ_j}{\sqrt{2}}\right)^2=iZ_j,\label{eq:step1}
\end{equation}
and that
\begin{align}
    \frac{1+iX_{j-1}X_{j}}{\sqrt{2}}\,  \prod_{m=j}^N \pqty{ iZ_m} \,\frac{1+iX_{j-1}X_{j}}{\sqrt{2}} 
    =\prod_{m=j}^N  \pqty{ iZ_m } \,\frac{1-iX_{j-1}X_{j}}{\sqrt{2}} \frac{1+iX_{j-1}X_{j}}{\sqrt{2}}
    =\prod_{m=j}^NiZ_m.\label{eq:step2}
\end{align}
Starting from the central terms in the product in Eq.~\eqref{eq:product} and repeatedly applying Eqs.~\eqref{eq:step1} and \eqref{eq:step2}, we derive
\begin{equation}
    S^2= i^N e^{-i\pi N/2}\prod_{m=1}^N Z_m=Q.
\end{equation}

We thus conclude that the Ising dual-reflection interface enjoys an enhanced $\mathbb{Z}_4$ symmetry\footnote{In contrast, in the ordinary transverse field Ising model, the Ising and reflection transformation constitute independent symmetries, resulting in the full symmetry group being $\mathbb{Z}_2\times \mathbb{Z}_2 $.} , generated by the $S$ operator. Therefore, it is possible to define a basis of energy eigenstates that carry a quantum number in the set $\{1, i ,-1, -i \}$ under $S\,.$ Since $S^2=Q$, the states with the quantum numbers $i$ or $-i$ have odd Ising charge ($Q=-1$), whereas the states with quantum numbers $-1$ or $1$ have even Ising charge ($Q=+1$). Similarly, we can classify operators according to how they change the $\mathbb{Z}_4$ charge of quantum states.

Finally, we note that ordinary symmetries generally preserve locality of operators. Self-duality transformations, on the other hand, although they commute with the Hamiltonian, do not necessarily map all local operators to local operators. Because $S$ acts non-locally on some local operators, see Eq.~\eqref{KW1}, it is not an ordinary symmetry but a self-duality. More precisely, since $S^4 = 1$, it is a self-duality of order four, also known as self-quadrality. To simplify our presentation, we refer to $S$ as a symmetry throughout this paper.

\subsection{Non-invertible symmetry in the closed chain}
Consider now a closed chain of $N$ sites, where we periodically identify the spins at the sites $j$ and $N+j$. In this setup, we introduce a second interface connecting the spins $j=N$ and $j=1\,,$ whereby we obtain the closed chain Hamiltonian
\begin{align} \label{H_spin_closed}
	\hop^{\circ} = \hop - J_0^\prime \, (\op{Z}{1} + \op{X}{N}\op{X}{1} )\,.
\end{align}
The new interface Hamiltonian transforms under the operation $S$ as
\begin{align}
	R \op{U}{KW} \, (\op{Z}{1} + \op{X}{N}\op{X}{1}) \, \op{U}{KW}^{-1} R^{-1}
    &= R \, (Q\op{X}{1}\op{X}{N} + Q\op{Z}{N} ) \, R^{-1}  \nonumber\\
	&= Q\op{X}{N}\op{X}{1} + Q\op{Z}{1} \,,
\end{align}
and thus breaks the $S$ symmetry. Note, however, that in the even Ising parity sector ($Q=+1$), $S$ actually acts as a symmetry of the total Hamiltonian $\hop^{\circ},$ for arbitrary coupling $J_0^\prime$ at the second interface. The symmetry operator appropriate for the closed chain must consequently project onto the even Ising parity sector and is therefore a non-invertible operator,
\begin{align} \label{S_0}
	S^{\circ} = S\,\frac{1 + Q}{2}\,,\qquad \bqty{S^{\circ},\, \hop^{\circ}} = 0\,,
\end{align}
which squares to the projector, $(S^{\circ})^2=(1+Q)/2$. This non-invertible symmetry is valid even away from criticality, in contrast to the well established Kramers-Wannier non-invertible symmetry in the isotropic transverse field Ising model.

\section{Fermionic description of the dual-reflection interface and its symmetry} \label{sec:ferm}

It has been appreciated for a long time that the quantum Ising chain in a transverse field can be solved by applying a Jordan-Wigner transformation, which transforms the Hamiltonian into a quadratic fermion model of a superconductor \cite{lieb1961two}. Here, we apply this transformation to the spin model of the dual-reflection interface and elucidate the action of the $\mathbb{Z}_4$ symmetry in the fermion formulation.

\subsection{Jordan-Wigner transformation}

The Jordan-Wigner transformation provides a non-local mapping between Hilbert spaces and operators of a spin $1/2$ chain on one side, and a single-component fermion chain on the other side. Specifically, for an open chain the local Pauli operators acting in the spin $1/2$ Hilbert space map onto
\begin{equation}
\begin{split} \label{eq:JW_X}
	\op{X}{j} \,&\rightarrow\, (-1)^{\sum_{k<j}\op{n}{k}}\, \pqty{\hc{c}{j} + \op{c}{j}}\,\ ,  \\
	\op{Z}{j} \, &\rightarrow\, (-1)^{\op{n}{j}}\,. 
\end{split}
\end{equation}
\vspace{-3pt}Here, the operators $ \op{c}{j} $ and $ \hc{c}{j} $ obey the canonical fermionic anticommutation relations, their occupation number operator is $  \op{n}{j}= \hc{c}{j}\op{c}{j}\,,$ and the local fermion parity operator is $(-1)^{\op{n}{j}} = 1-2\op{n}{j}$. Except for sites $j$ adjacent to the edge $j=1$, any individual $\op{X}{j}$ is mapped onto an operator with a non-local Jordan-Wigner parity string. Under the mapping, the Ising symmetry operator $Q=\prod_j \op{Z}{j}$ becomes the fermion parity operator $\mathcal{P}=\prod_j (-1)^{\op{n}{j}}$ which generates the global $\mathbb{Z}_2$ symmetry in the fermion formulation.

It is convenient to introduce a pair of Majorana operators for each complex fermion mode,
\begin{equation}
\begin{split}\label{eq:eta_def}
	\op{\eta}{2j-1} &= \hc{c}{j} + \op{c}{j} \,, \\
	\op{\eta}{2j} &= i\,\pqty{\hc{c}{j} - \op{c}{j}} \,, 
\end{split}
\end{equation}
satisfying the anticommutation relation $\{\eta_a, \eta_b\}=2\delta_{a,b}\,.$
In terms of Majorana operators, the Jordan-Wigner transformation reads
\begin{equation}\begin{split}\label{eq:JW_Maj}
	\op{X}{j}\op{X}{j+1} \,&\rightarrow\, -i \, \op{\eta}{2j}\op{\eta}{2j+1}\ ,\\
	\op{Z}{j} \, &\rightarrow\, -i\, \op{\eta}{2j-1}\op{\eta}{2j}\,.
\end{split}
\end{equation}
Clearly, the Hamiltonian of the transverse field Ising chain becomes quadratic in Majoranas (for any choice of position-dependent Ising and transverse field couplings). In particular, the Hamiltonian \eqref{H_int} of the open chain with dual-reflection interface maps on\footnote{We use calligraphic and ordinary fonts for operators in the fermion and spin language respectively.} (see Fig.~\ref{fig:Majorana_chain})
\begin{align} \label{H_f}
	\mathcal{H} = i \sum_{j=1}^{N-1} J_j\, \op{\eta}{2j}\op{\eta}{2j+1} + i  \sum_{j=2}^N h_j \,\op{\eta}{2j-1}\op{\eta}{2j}\,,
\end{align}
with the couplings
\begin{equation}
\begin{split}
	&J_j = \begin{cases}
		J & j = 1, \,\dots,\, n-1 \\
		J_0 & j = n\\
		h & j = n+1,\,\dots,\,N-1
	\end{cases}\,, \\
	&h_j = \begin{cases}
		h & j = 2, \,\dots,\, n \\
		J_0 & j = n+1\\
		J & j = n+2,\,\dots,\,N
	\end{cases}\,,
\end{split}
\end{equation}
where $n=N/2\,.$

\subsection{$\mathbb{Z}_4$ symmetry}\label{sec:ferm_Z4}

Here, we will elucidate the $\mathbb{Z}_4$ symmetry of the Ising dual-reflection interface in the fermion formulation.

First, as has already been shown in \cite{seiberg2024non} and as detailed in Appendix~\ref{appB}, the Kramers-Wannier unitary \eqref{KW} transforms under the Jordan-Wigner mapping to the Majorana sequential circuit \cite{PhysRevB.109.075116}
\begin{equation}
\mathcal{U}_{KW} = e^{-2\pi i N/8}\, \prod_{a=2N-1}^{1} \frac{1+\op{\eta}{a}\op{\eta}{a+1}}{\sqrt{2}}\,,
\end{equation}
which acts on individual Majoranas as
\begin{equation}\begin{split} \label{eq:KW_M}
        \mathcal{U}_{KW}\,\op{\eta}{1}\,\mathcal{U}_{KW}^\dagger &= - \op{\eta}{2N}\, , \\
	\mathcal{U}_{KW}\,\op{\eta}{a}\,\mathcal{U}_{KW}^\dagger &=  \op{\eta}{a-1}\,
\end{split}
\end{equation}
for $a=2,3,\dots, 2N \,$.
On a closed Majorana chain  made of $2N$ sites (with antiperiodic boundary condition $\op{\eta}{2N+1}\equiv -\op{\eta}{1} $), this transformation corresponds to an elementary (counter-clockwise) translation \cite{seiberg2024non}. In contrast,  it loosely speaking acts as ``half-elementary translation'' on the original spin $1/2$ chain.

Second, as derived in Appendix~\ref{appB}, the spatial reflection symmetry that swaps the spin sites $j\leftrightarrow N+1-j$ acts on individual Majorana fermions as
\begin{equation}\begin{split} \label{eq:KW_r}
	\mathcal{R}\, \op{\eta}{2j-1} \, \mathcal{R}^{-1} &=  i \mathcal{P} \, \op{\eta}{\widehat{2j-1}}\,,\\
	\mathcal{R}\,\op{\eta}{2j} \,\mathcal{R}^{-1} &= \ -i \mathcal{P} \, \op{\eta}{\widehat{2j}}\, ,
\end{split}\end{equation}
where $\widehat{a}=2N+1-a$ denotes the reflection of $a$ with respect to the link between the sites $N$ and $N+1$.
By explicit calculation, we find $\mathcal{R}^2 \op{\eta}{a} \mathcal{R}^{-2}=\op{\eta}{a}$.

\begin{figure}
    \centering
    \begin{tikzpicture}[scale=0.9]
		\tikzstyle{every node}=[font=\fontsize{14}{14}\selectfont]
		
		\foreach \x in { -4,1}{
			\draw[orange] (\x+0.5, 0.25) node[anchor=south]{$J$};
			\draw[ orange, line width=3pt] (\x,0) -- (\x+1,0);
			
		}
		
		\draw[anthracite] (-1.5, 0.25) node[anchor=south]{$J_0$};
		\draw[anthracite] (-0.5, 0.25) node[anchor=south]{$J_0$};
		\draw[ line width=3pt, anthracite] (-2,0) -- (0,0);
		\foreach \x in {-3,0}{
			\draw[green] (\x+0.5, 0.25) node[anchor=south]{$h$};
			\draw[ green, line width=3pt] (\x,0) -- (\x+1,0);
		}
		
		\foreach \x in {-4,...,2}
		{        
			\filldraw[fill=white] (\x,0) circle (0.1);
		}
        \draw (-1,-0.7) node[anchor=south, font=\fontsize{11}{11}\selectfont]{$N$$+1$};
        \draw (-3,-0.7) node[anchor=south, font=\fontsize{11}{11}\selectfont]{$a$};
        \draw (1,-0.7) node[anchor=south, font=\fontsize{11}{11}\selectfont]{$\widetilde{a}$};
        
		\begin{scope}[shift={(-2,1)}, rotate=180]
    		\draw[<->, thick, blue] (-1.5,-0.25)..controls (-1,-0.5) ..(-0.5,-0.25);
    		\draw[<->, thick, blue] (-2.5,-0.25)..controls (-1,-0.75) ..(0.5,-0.25);
    		\draw[<->, thick, blue] (-3.5,-0.25)..controls (-1,-1) ..(1.5,-0.25);
    		\draw[blue] (-1,-1.5) node[anchor=north]
        {$\mathcal{S}=\mathcal{R}\mathcal{U}_{KW}$};
        \end{scope}
  
		\draw (-4.25,0) node[anchor=east]{...};
		\draw (2.25,0) node[anchor=west]{...};

        \begin{scope}[shift={(6.5,0.5)}, scale=1]
        \draw[line width=3pt] (-1,0)--(0,0);
			\foreach \x in {-1,0}
    		{        
    			\filldraw[fill=white] (\x,0) circle (0.1);
    		}
        \draw[] (-1,-0.1) node[anchor=north,font=\fontsize{11}{11}\selectfont]{$a$};
        \draw[] (0,0) node[anchor=north,font=\fontsize{11}{11}\selectfont]{${a+1}$};
        \draw[] (0.5,0) node[anchor=west]{$i\eta_a\eta_{a+1}$};
		\end{scope}
        \draw[] (5.2,-0) rectangle (9,1);
	\end{tikzpicture}
    \caption{Schematic action of the Kramers-Wannier transformation combined with spatial reflection on the Ising dual-reflection interface in its fermion formulation: The empty circles represent Majorana sites, and the coupling strengths between them are colour-coded.  The transformation of an individual Majorana operator $\eta_a$ under $\mathcal{S}=\mathcal{R} \mathcal{U}_{KW}$ amounts to a reflection about the Majorana site $N+1$, which exchanges $\eta_a$ and $\eta_{\widetilde{a}}\,,$ and multiplication by $\pm i \mathcal{P}$.}
    \label{fig:Majorana_chain}
\end{figure}

Combining the operations $ \mathcal{R} $ and $ \mathcal{U}_{KW} $ yields the $\mathbb{Z}_4$ symmetry $ \mathcal{S}=\mathcal{R} \mathcal{U}_{KW}\,,$ which acts on the Majorana fermions as
\begin{equation}\begin{split} \label{eq:S_eta}
        \mathcal{S}\,\op{\eta}{1}\,\mathcal{S}^{-1} &=i \mathcal{P} \op{\eta}{1}\, , \\
	\mathcal{S}\,\op{\eta}{2j-1}\,\mathcal{S}^{-1} &= - i \mathcal{P} \, \op{\eta}{\widetilde{2j-1}} \qquad\text{for } j=2,3,\dots, N \,,\\
        \mathcal{S}\,\op{\eta}{2j}\,\mathcal{S}^{-1} &= i \mathcal{P} \, \op{\eta}{\widetilde{2j}}\, ,
\end{split}\end{equation}
where we introduced the notation $\widetilde{a}=2N+2-a$ to denote the reflection of $a$ with respect to the site $N+1$.
In a closed chain, the first two equations combine to $\mathcal{S}\,\op{\eta}{2j-1}\,\mathcal{S}^{-1} = - i \mathcal{P} \, \op{\eta}{\widetilde{2j-1}}$\vspace{-1pt} provided we impose the antiperiodic boundary condition $\op{\eta}{2N+1}\equiv -\op{\eta}{1} $. 
In the fermion representation, the $\mathcal{S}$ symmetry thus has an apparent manifestation: for the chain with $2N$ Majorana sites it involves a spatial reflection $a \leftrightarrow \widetilde{a}=2N+2-a$ with respect to the Majorana site $a=N+1\,,$ see Fig.~\ref{fig:Majorana_chain}. When conjugating a product of an odd number of Majorana operators with $\mathcal{S}$, the result in addition depends on the fermion parity. Clearly, this transformation differs from the ordinary spatial reflection $\mathcal{R}$ that in the fermion formulation swaps Majorana sites $a\leftrightarrow \widehat{a}=2N+1-a$, i.e., reflects not about a site but with respect to the link connecting the Majorana sites $N$ and $N+1\,.$\footnote{In contrast to the spatial reflection defined in spin language, the dual reflection implemented by $\mathcal{S}$ cannot be formulated well in terms of pairs of Majorana fermions $\eta_{2j-1},\,\eta_{2j}$ associated with spin site $j$, since it reflects the two Majoranas to different, adjacent spin sites.} 

The two Majorana fermions $\op{\eta}{1}$ and $\op{\eta}{N+1}$ are special as they map on themselves under $\mathcal{S}$ (up to a parity-dependent factor).
It is convenient to define $2N$ new Majorana operators
\begin{equation}\label{eq:xi}
    \xi_a^{\pm}=\frac{1}{\sqrt{2}}\,\bqty{\eta_a \pm (-1)^a\, \eta_{\widetilde a}}
\end{equation}
for $a=2,\dots,N\,$, and
\begin{equation}\label{eq:xi_special}
    \xi_1^+=\eta_1, \qquad \xi_{N+1}^-=\eta_{N+1},
\end{equation}
such that they all transform as
\begin{equation} \label{eq:S_xi}
    \mathcal{S}\xi_a^{\pm}\mathcal{S}^{-1}=\pm i \mathcal{P}\,\xi_a^{\pm}\,,
\end{equation}
and obey Majorana anticommutation relations $\Bqty*{\xi_a^p,\, \xi_b^q} = 2\delta_{a,\,b} \,\delta_{q,\,p}\,,$ with $q,\,p = \pm\,.$
The Hamiltonian of the open chain \eqref{H_f} is conveniently expressed in terms of these Majorana modes as
\begin{equation} \label{eq:H_Majorana_open}
\begin{split}
    \mathcal{H} = &\,ih \sum_{j=2}^{n} \pqty{ \xi_{2j-1}^+ \xi_{2j}^+ + \xi_{2j-1}^- \xi_{2j}^-} + \sqrt{2} iJ_0 \,\xi_N^- \xi_{N+1}^- + iJ \sum_{j=1}^{n-1} \pqty{\xi_{2j}^+ \xi_{2j+1}^+ + \xi_{2j}^- \xi_{2j+1}^-}  \,.
\end{split}
\end{equation}
It is straightforward to check that the open-chain Hamiltonian written in this way commutes with the operator $\mathcal{S}$. Generically, only Majorana modes of the same sign can be coupled quadratically in an $\mathcal{S}$-symmetric manner.\footnote{If we include interactions in the Hamiltonian, for example a local $\mathcal S$-invariant term like $\eta_2 \eta_3 \eta_4 \eta_5+\eta_{\widetilde{5}} \eta_{\widetilde{4}} \eta_{\widetilde{3}} \eta_{\widetilde{2}}
=\frac{1}{2}\,
\big(\xi_2^+\xi_3^+\xi_4^+\xi_5^+
    +\xi_2^+\xi_3^-\xi_4^-\xi_5^+
    +\xi_2^-\xi_3^-\xi_4^+\xi_5^+
    +\xi_2^-\xi_3^+\xi_4^-\xi_5^+
    +\xi_2^-\xi_3^+\xi_4^+\xi_5^-
    +\xi_2^+\xi_3^-\xi_4^+\xi_5^-
    +\xi_2^+\xi_3^+\xi_4^-\xi_5^-$ $
    +\,\xi_2^-\xi_3^-\xi_4^-\xi_5^-\big)$, we can couple the two chains without breaking the $\mathbb{Z}_4$ symmetry. This decoupling is therefore not a general property of $\mathcal S$-invariant Hamiltonians.}
As illustrated in Fig.~\ref{fig:double_chain}, in the new representation we obtain two decoupled Majorana chains governed by the Hamiltonians
\begin{equation} \label{eq:H_two_Majorana}
\begin{split}
    \mathcal{H}^+ =&\, ih \sum_{j=2}^{n}  \xi_{2j-1}^+ \xi_{2j}^+ + iJ \sum_{j=1}^{n-1} \xi_{2j}^+ \xi_{2j+1}^+ \, , \\
    \mathcal{H}^- =\,& ih \sum_{j=2}^{n} \xi_{2j-1}^- \xi_{2j}^- + iJ \sum_{j=1}^{n-1}  \xi_{2j}^- \xi_{2j+1}^- + \sqrt{2} iJ_0 \,\xi_N^- \xi_{N+1}^- \, ,
\end{split}
\end{equation}
which commute with each other, $\bqty{\mathcal{H}^+,\mathcal{H}^-}=0\,$. The total Hamiltonian is $\mathcal H =\mathcal{H}^+ + \mathcal{H}^-$.
The fermion parity associated with each chain is conserved separately. Locality (in terms of the original Majoranas $\eta_a$) ensures that, firstly, $\xi_1^-=\eta_1$ decouples from the Hamiltonian, and that, secondly, corresponding $\xi^+$ and $\xi^-$ bilinears appear with the same coupling strength in the two sums in Eq.~\eqref{eq:H_Majorana_open}.

\begin{figure}
    \centering
    \begin{tikzpicture}[scale=0.70]
		\tikzstyle{every node}=[font=\fontsize{14}{14}\selectfont]
        \draw[white] (-11,1) rectangle (9,-1); 
		\foreach \x in { -6,-2,2}{
			\draw[orange, line width=3pt] (\x,0.7) -- (\x+2,0.7);	
		}
		
		\foreach \x in { -4,0,4}{
			\draw[green, line width=3pt] (\x,0.7) -- (\x+2,0.7);	
		}
		
		\draw[anthracite, line width=3pt] (6,0.7)..controls (7.5,0.7) and (8,0.7) ..(8,0);
		\draw[anthracite, line width=3pt] (6,-0.7)..controls (7.5,-0.7) and (8,-0.7) ..(8,0);
		\filldraw[fill=white] (8,0) circle (0.15);
		
		\foreach \x in { -6,-2,2}{
			\draw[orange, line width=3pt] (\x,-0.7) -- (\x+2,-0.7);	
		}
		
		\foreach \x in { -4,0,4}{
			\draw[green, line width=3pt] (\x,-0.7) -- (\x+2,-0.7);	
		}
		
		\foreach \x in {-6,-4,...,6}
		{        
			\filldraw[fill=white] (\x,0.7) circle (0.15);
			\filldraw[fill=white] (\x,-0.7) circle (0.15);
		}
		\filldraw[fill=white] (-8,0.7) circle (0.15);
		
		\draw (-1,0.7) node[fill=white]{...};
		\draw (1,0.7) node[fill=white]{...};
		
		\draw (-1,-0.7) node[fill=white]{...};
		\draw (1,-0.7) node[fill=white]{...};
		
		\draw (-8,1) node[anchor=south]{$\eta_1$};
		\draw (-6,1) node[anchor=south]{$\eta_2$};
		\draw (0,1) node[anchor=south]{$\eta_{a}$};
		\draw (8,0) node[anchor=east]{$\eta_{N+1}$};
		\draw (0,-1) node[anchor=north]{$\eta_{\widetilde{a}}$};
		\draw (-6,-1) node[anchor=north]{$\eta_{2N}$};
		
		\draw[blue, line width=1pt] (0,0) ellipse (0.4 and 1);
		
		\draw[->, thick, blue] (0.5,-0)..controls (1.75,-0.5) and (1.75,-2)..(0.5,-4);
	    \node[blue, font=\fontsize{11}{11}, anchor=west] at (1.6,-1.7) {$\xi^{\pm}_a = \frac{1}{\sqrt{2}}\,\big(\eta_a \pm (-1)^a \eta_{\widetilde{a}}\big)$};
		\begin{scope}[shift={(0,-4)}]
			\foreach \x in { -6,-2,2}{
				\draw[orange, line width=3pt] (\x,0.7) -- (\x+2,0.7);	
			}
			
			\foreach \x in { -4,0,4}{
				\draw[green, line width=3pt] (\x,0.7) -- (\x+2,0.7);	
			}
			
			\foreach \x in { -6,-2,2}{
				\draw[orange, line width=3pt] (\x,-0.7) -- (\x+2,-0.7);	
			}
			
			\draw[anthracite, line width=3pt] (6,-0.7) -- (8,-0.7);	
			
			\foreach \x in { -4,0,4}{
				\draw[green, line width=3pt] (\x,-0.7) -- (\x+2,-0.7);	
			}
			
			\foreach \x in {-6,-4,...,6}
			{        
				\filldraw[fill=white] (\x,0.7) circle (0.2);
				\draw[line width=1.2pt] (\x,0.55)--(\x,0.85);
				\draw[line width=1.2pt] (\x-0.15,0.7)--(\x+0.15,0.7);
				\filldraw[fill=white] (\x,-0.7) circle (0.2);
				\draw[line width=1.2pt] (\x-0.15,-0.7)--(\x+0.15,-0.7);
			}
			\filldraw[fill=white] (8,-0.7) circle (0.2);
            \draw[line width=1.2pt] (7.85,-0.7)--(8.15,-0.7);
			\filldraw[fill=white] (-8,0.7) circle (0.2);
            \draw[line width=1.2pt] (-8,0.55)--(-8,0.85);
			\draw[line width=1.2pt] (-7.85,0.7)--(-8.15,0.7);
			
			\draw (-1,0.7) node[fill=white]{...};
			\draw (1,0.7) node[fill=white]{...};
			
			\draw (-1,-0.7) node[fill=white]{...};
			\draw (1,-0.7) node[fill=white]{...};
			
			\draw (-8,0) node[font=\fontsize{11}{11}]{$1$};
			\draw (-6,0) node[font=\fontsize{11}{11}]{$2$};
            \draw (6,0) node[font=\fontsize{11}{11}]{$N$};
			\draw (8,0) node[font=\fontsize{11}{11}]{$N$$+$$1$};
			\draw (0,0) node[font=\fontsize{11}{11}]{$a$};
			
			\draw[blue, line width=1pt] (0,0.7) circle (0.4);
			\draw[blue, line width=1pt] (0,-0.7) circle (0.4);

            \node[] at (-9.75,0.8) {$\mathcal{H}^+$};
            \node[] at (-9.75,-0.6) {$\mathcal{H}^-$};
		\end{scope}
\begin{scope}[shift={(-10.5,1.5)}]
	\draw[] (-0.2,0.4) rectangle (1.8,-1.6);
	\draw[orange, line width=3pt] (0,0) -- (1,0);
	\draw[] (1,0) node[anchor=west, font=\fontsize{11}{11}]{$J$};
	\draw[green, line width=3pt] (0,-0.6) -- (1,-0.6);
	\draw[] (1,-0.6) node[anchor=west, font=\fontsize{11}{11}]{$h$};
	\draw[anthracite, line width=3pt] (0,-1.2) -- (1,-1.2);
	\draw[] (1,-1.2) node[anchor=west, font=\fontsize{11}{11}]{$J_0$};
\end{scope}	
	\end{tikzpicture}
\caption{Decomposition of the open quadratic Majorana chain with ${\mathcal{S}}$-symmetry: The fermion model \eqref{H_f}, illustrated in the upper part of the figure, splits into two Majorana chains with the Hamiltonians $\mathcal{H^+}$ and $ \mathcal{H}^-$ which decouple from each other (see Eq.~\eqref{eq:H_two_Majorana}). To reveal this, we combine the Majoranas $\eta_a$ and $\eta_{\widetilde{a}}$  into Majoranas $\xi_a^\pm$ which obey $\mathcal{S}\xi_a^\pm \mathcal{S}^{-1}=\pm i\mathcal{P}\, \xi_a^\pm $. They form the two separate chains depicted in the lower part. The Majorana $\eta_1=\xi^+_1$ belongs to the $\xi^+$ chain, but is not connected to other Majoranas in the open geometry due to locality; the interface Majorana $\eta_{N+1}=\xi_{N+1}^-$ is part of the $\xi^-$ chain.}
    \label{fig:double_chain}
\end{figure}

Now, we will demonstrate that $ \mathcal{S}^2=\mathcal{P} $ in the fermion formulation.\footnote{Various other models with symmetries squaring to the fermion parity can be found in the literature. For example, a reflection symmetry of the homogeneous (closed) Kitaev chain squares to the fermion parity $\mathcal{P}$\cite{shapourian2017many, shiozaki2017many1}. A $\mathbb{Z}_4$ symmetry squaring to the fermion parity arises moreover in the context of $\mathbb{Z}_4$ parafermions \cite{calzona2018parafermions, chew2018parafermions, mazza2018parafermions}.} First, we notice that
\begin{equation}
	\mathcal{R}\,\mathcal{U}_{KW}\, \mathcal{R} = \, e^{-2\pi i N/8}\,\frac{1+\op{\eta}{1}\op{\eta}{2}}{\sqrt{2}} \frac{1+\op{\eta}{2}\op{\eta}{3}}{\sqrt{2}}\,\dots\, \frac{1+\op{\eta}{2N-1}\op{\eta}{2N}}{\sqrt{2}}\,.
\end{equation}
Multiplying by $\mathcal{U}_{KW}$ from the right  and following the steps in the previous proof of the identity $ S^2 = Q $, we find
\begin{align} 
	\mathcal{S}^2 = (-i)^N\,\prod_{a=1}^{2N} \op{\eta}{a} = \prod_{j=1}^N \pqty{-i \op{\eta}{2j-1} \op{\eta}{2j} }= (-1)^{\sum_{j=1}^N \op{n}{j}}=\mathcal{P}\,.
\end{align}
Consequently, in contrast to the ordinary reflection $\mathcal{R}$ (which squares to unity when acting on a Majorana fermion, $\mathcal{R}^2 \op{\eta}{a} \mathcal{R}^{-2}=\op{\eta}{a}$) the symmetry $\mathcal{S}$ squares to minus unity, i.e., we find $\mathcal{S}^2 \op{\eta}{a} \mathcal{S}^{-2}=-\op{\eta}{a}$.

Next, we will investigate the fermion dual of the closed spin chain (with periodic boundary conditions for spins) governed by the Hamiltonian \eqref{H_spin_closed} that commutes with the non-invertible symmetry \eqref{S_0}. Given periodic boundary conditions for the spin operators in the Ising model, the non-local Jordan-Wigner string leads to a dependence of the boundary conditions for the Majorana fermions on the value of the fermion parity $\mathcal{P}$, see e.g. \cite{shankar2017quantum, mbeng2024quantum}. Namely, one must use antiperiodic (periodic) boundary conditions for the Majorana fermions in the even $\mathcal{P}=1$ (odd $\mathcal{P}=-1$) parity sector. 

Since the symmetry $S$ of the dual-reflection interface in the closed geometry is valid only in the even Ising parity sector ($Q=1$), we must impose antiperiodic boundary conditions on the Majorana fermions, $\eta_{2N+1}=-\eta_1\,.$ It is straightforward to check explicitly that the Hamiltonian
\begin{equation}\label{H_Majorana_closed}
	\mathcal{H}^{\circ} = \mathcal{H} + i J_0^\prime\, (- \op{\eta}{2N}\op{\eta}{1}+\op{\eta}{1}\op{\eta}{2} ) = \mathcal{H} + \sqrt{2} i J_0^\prime\, \xi_1^+ \xi_2^+ \,,
\end{equation}
containing an end-to-end coupling of strength $J_0^\prime\,,$ commutes with the $\mathbb{Z}_4$ symmetry generator $\mathcal{S}$. Closing the spin chain corresponds to coupling the Majorana $\xi_1^+=\eta_1$ to the $\xi^+$ Majorana chain, see Fig.~\ref{fig:two_Majorana_closed}. Clearly, $\mathcal{H}^{\circ}$ is the Jordan-Wigner transformation of the spin Hamiltonian $ H^{\circ}$ in the $Q=1$ symmetry sector. Note, however, that in the fermion theory, even for closed boundary conditions, the symmetry $\mathcal{S}$ is valid in the full Hilbert space (including the $\mathcal P=-1$ sector) and thus generates a $\mathbb{Z}_4$ group.

\section{Majorana strong zero modes } \label{sec:SMZM}
It is a well-known fact that an open Kitaev chain in the topological phase supports two Majorana modes that are spatially localized near the two edges and ensure a robust ground state degeneracy \cite{kitaev2001unpaired}. More generally, these are examples of the so-called strong zero modes introduced in \cite{alicea2016topological, fendley2016strong}. For the purpose of this paper, we define a strong zero mode $\Psi$ as a spatially localized operator that
\begin{itemize}
\item commutes with the Hamiltonian up to corrections which are exponentially small in the system size,
\item anticommutes with an exact $\mathbb{Z}_2$ symmetry of the model,
\item satisfies $\Psi^2=1$.
\end{itemize}
By acting on an energy eigenstate, $\Psi$ generates an orthogonal state of the same energy (up to a small error scaling exponentially with the system size) with a toggled eigenvalue of the $\mathbb{Z}_2$ symmetry. As a result, the whole energy spectrum is at least twofold degenerate. 
Another physical consequence of the existence of the strong zero mode is that any operator that overlaps strongly with $\Psi$ must evolve very slowly in time \cite{kemp2017long}.

In this section, using the fermion formulation of Sec.~\ref{sec:ferm},  we investigate the fate of Majorana strong zero modes in our model and examine the role played by the $\mathbb{Z}_4$ symmetry.  While we justify and study all strong zero modes in the main text, their detailed construction using the Bogoliubov-de Gennes equations (introduced in Appendix~\ref{app:BdG}) is provided in Appendix~\ref{app:SZMs}.

\subsection{Open chain}
In the following, we discuss Majorana strong zero modes in the open chain, where we have to distinguish the two regimes $J>h$ and $J<h$ in general. We first demonstrate that the $\mathbb{Z}_4$ symmetry $\mathcal{S}\,$, along with locality, entails a pair of exact Majorana strong zero modes that strictly commute with the Hamiltonian, and anticommute with the fermion parity. This result holds for any choice of coupling constants and ensures an exact twofold degeneracy of all energy eigenstates in the open chain. It is easy to identify one of these exact Majorana strong zero modes. The remaining Majorana strong zero modes are constructed separately in the $J>h$ and $J<h$ regimes. Near the critical point $h\rightarrow J\,,$ these modes delocalize and can no longer be normalized in the thermodynamic limit. Finally, we outline how $\mathcal{S}$-preserving or $\mathcal{S}$-breaking perturbations affect the presence of the strong zero modes.

As discussed above, in the open chain the leftmost Majorana coupling must be set to zero ($h_1=0$) to obtain the $\mathcal{S}$-symmetric local Hamiltonian \eqref{H_f}.  One exact strong zero mode\footnote{In the spin formulation of Sec.~\ref{sec:spin}, the left-boundary-localized zero mode $\eta_1$ simply becomes the local spin operator $X_1$ which anticommutes with the Ising $\mathbb{Z}_2$ symmetry and strictly commutes with the Hamiltonian \eqref{H_int}.} is therefore the local Majorana fermion $\xi_1^+=\eta_1$, which clearly commutes with the Hamiltonian \eqref{H_f}, and anticommutes with the fermion parity $\mathcal{P}$. Importantly, the exactness of this strong zero mode is robust under adding arbitrary local interaction terms that respect the symmetry $\mathcal{S}$ to the quadratic model \eqref{H_f}.

\subsubsection{$J>h$ regime}\label{sec:open_chain_J>h}

In the case $J>h$, the $\xi^+$ chain in Fig.~\ref{fig:double_chain} is in the topological phase hosting a pair of Majorana edge modes. The left one, $\xi^+_1=\eta_1$, was discussed above. 
The second exact Majorana strong zero mode is localized around the dual-reflection interface and must be a superposition of $\xi^+_a$ Majoranas. It
can be constructed using the usual iterative procedure: First, we identify $\xi_{N}^{+} $ as an operator that commutes with the Hamiltonian to zeroth order in $h$,
\begin{align} \label{eq:com}
    \bqty{\mathcal{H},\, \xi_{N}^{+}} = 2ih\, \xi_{N-1}^{+}\,,
\end{align}
and make the ansatz
\begin{align} \label{eq:eta_an}
    \eta = \frac{1}{\mathcal{N}} \,\bqty{\xi^+_N + \mathcal{O}({\scriptstyle\frac{h}{J}}) }\,.
\end{align}
The observation
\begin{align}\label{eq:bulk_commutator_xi}
    \bqty{\mathcal{H},\, \xi_{2j}^{\pm}} = 2ih\, \xi_{2j-1}^{{\pm}} - 2iJ\, \xi_{2j+1}^{{\pm}}
\end{align}
for $j=2,\dots,N/2 \,$,
shows that we can cancel the first order term \eqref{eq:com} in the commutator of $\eta$ with $\mathcal H$ by adding $h/J\,\xi_{N-2}^+$ to $\eta$ in Eq.~\eqref{eq:eta_an}. We can now iterate this cancellation procedure until we reach $\xi_2^+$. Due to the vanishing coupling $h_1$, we find 
\begin{align}
    \bqty{\mathcal{H},\, \xi_{2}^{+}} = - 2iJ\, \xi_{3}^{{+}} \,,
\end{align}
which is proportional to what is required to compensate the finite contribution to the commutator $\bqty{\mathcal H,\,\eta}$ stemming from $\xi_4^+\,.$
In total, we discover an exact Majorana strong zero mode
\begin{align} \label{eq:eta}
    \eta = \frac{1}{\mathcal{N}}\sum_{j=1}^{n} \pqty{\frac{h}{J}}^{n-j}\,\xi_{2j}^+\,,\qquad \bqty{\mathcal{H},\, \eta}=0\,,
\end{align}
which is symmetrically and exponentially localized around the interface, and anticommutes with the fermion parity $\mathcal{P}$.
The normalization factor reads
\begin{align}
	\mathcal{N}^2 = J^2\,\frac{ 1-\pqty{\frac{h}{J}}^{{N}} }{ J^2-{h^2} }\,. 
\end{align}
From Eqs.~{\eqref{eq:xi_special}} and \eqref{eq:S_xi} it is evident that both Majorana zero modes $\eta_1 $ and $\eta$ transform alike under the $\mathcal{S}$ symmetry. They can be combined into a complex fermion operator $(\eta+i \eta_1)/2$ acting on a two-dimensional Hilbert space. This gives rise to a strict, twofold degeneracy of each level of the energy spectrum. In contrast to the Kitaev chain, the $\mathcal{S}$ symmetry together with locality suffices to ensure that this result is exact for an open chain and does not receive any finite-size corrections.

\subsubsection{$J<h$ regime}
In the regime $J<h$, instead of the exact interface strong zero mode we find an exact strong zero mode bound to both edges of the chain, symmetrically. To leading order, the exact edge-strong zero mode is given by
\begin{align}
    \eta = \frac{1}{\mathcal{N}}\,\bqty{\xi_2^+ + \mathcal{O}({\scriptstyle\frac{J}{h}})}\,.
\end{align}
The same steps as above lead us to the operator
\begin{align} \label{eq:eta_n}
    \eta =  \frac{1}{\mathcal{N}}\sum_{j=1}^{n} \pqty{\frac{J}{h}}^{j-1}\,\xi_{2j}^+\,,
\end{align}
which obeys
\begin{align}
     \bqty{\mathcal{H},\, \eta}=0\,, \qquad \mathcal{S}\,\eta\,\mathcal{S}^{-1} = i\mathcal{P}\, \eta\,.
\end{align}
It is normalized by
\begin{align}\label{eq:eta_edge_norm}
     \mathcal{N}^2 = h^2\,\frac{ 1-\pqty{\frac{J}{h}}^N }{h^2-J^2} \,.
\end{align}
The exact Majorana modes $\eta$ and $\eta_1$ together form a complex fermion causing a strict twofold degeneracy of the energy spectrum.\footnote{Equations \eqref{eq:eta} and \eqref{eq:eta_n} are different expansions of one and the same operator. The two expressions make it apparent that the localization regions of the corresponding zero mode change as we move between the two regimes $J>h$ and $h>J$.} In the illustration in Fig.~\ref{fig:double_chain}, they are both localized near the left boundary of the $\xi^+$ chain.

Moreover, for $J<h$, the $\xi^-$ chain in Fig.~\ref{fig:double_chain} is in the topological phase and two Majorana strong zero modes made of $\xi^-$ fermions must exist. A strong zero mode localized around the interface is given by
\begin{align} \label{etapr}
    \eta^\prime =  \frac{1}{\mathcal{N}^\prime}\bqty{\frac{1}{\sqrt{2}}\,{\xi_{N+1}^-} + \frac{J_0}{h} \sum_{j=0}^{n-2} \pqty{\frac{J}{h}}^{j}\,\xi_{N-(2j+1)}^{-}}\,,
\end{align}
with the normalization
\begin{align}
    {\mathcal{N}^\prime}^2 = \frac{1}{2} + J_0^2\,\frac{ 1 - \pqty{\frac{J}{h}}^{N-2} }{h^2 - J^2} \,.
\end{align}
It commutes with the Hamiltonian up to a small term decaying exponentially with system size,\footnote{Clearly, if we decouple the left and right parts of the chain by setting $J_0=0$, the Majorana strong zero mode $\eta'=\eta_{N+1}$ becomes exact, i.e., $\bqty{\mathcal{H},\, \eta^\prime} = 0$. }
\begin{align}
     \bqty{\mathcal{H},\, \eta^\prime} = \frac{1}{\mathcal{N}^\prime}\,2iJ_0\, e^{-\pqty{n-1}\,\lambda^{-1}}\, \xi_2^-\,,
\end{align}
with $\lambda^{-1} = \ln{\pqty{\frac{h}{J}}}\,$. Another Majorana strong zero mode, which is localized around the edges, can be obtained from $\eta$ by replacing all $\xi_{2j}^+$ with $\xi_{2j}^-\,,$
\begin{equation}\label{eq:eta_nn}
    \eta^{\prime\prime} =  \frac{1}{\mathcal{N}}\sum_{j=1}^{n} \pqty{\frac{J}{h}}^{j-1}\,\xi_{2j}^-\,,
\end{equation}
Its normalization is given in Eq.~\eqref{eq:eta_edge_norm}. Due to the local nature of the couplings, the commutator with the Hamiltonian changes only around the interface, where $\eta^{\prime\prime}$ vanishes in the thermodynamic limit,
\begin{equation}
     \bqty{\mathcal{H},\, \eta^{\prime\prime}} = - \frac{1}{\mathcal{N}}\,2\sqrt{2}i J_0\, e^{-\pqty{n-1}\,\lambda^{-1}}\, \xi_{N+1}^-\,,
\end{equation}
with $\lambda^{-1} = \ln{\pqty{\frac{h}{J}}}\,$.
Thus, $\eta^{\prime\prime}$ is a valid strong zero mode. Since the modes $\eta'$ and $\eta''$ both reside on the $\xi^-$ chain in Fig.~\ref{fig:double_chain}, they have the same transformation property (see Eq.~\eqref{eq:S_xi}),
\begin{equation}
    \mathcal{S}\,\eta^{\prime}\,\mathcal{S}^{-1} = -i\mathcal{P}\, \eta^{\prime}\,,
    \quad
    \mathcal{S}\,\eta^{\prime\prime}\,\mathcal{S}^{-1} = -i\mathcal{P}\, \eta^{\prime\prime}\,.
\end{equation}
One can combine these two modes into a complex fermion $(\eta'+i \eta'')/2$.

In the following, we will moreover use that the modes $\eta$ and $\eta''$ can be combined to obtain the two modes $\eta_L$ and $\eta_R$ localized on the left and right edge of the chain respectively,
\begin{equation} \label{eq:eta_L}
    \eta_L=\frac{1}{\sqrt{2}}\pqty{\eta+\eta^{\prime\prime}}=\frac{1}{\mathcal N}\sum_{j=1}^{n}\left(\frac{J}{h}\right)^{j-1} \eta_{2j},
\end{equation}
\begin{equation} \label{eq:eta_R}
    \eta_R=\frac{1}{\sqrt{2}}\pqty{\eta-\eta^{\prime\prime}}=\frac{1}{\mathcal N}\sum_{j=1}^{n}\left(\frac{J}{h}\right)^{j-1} \eta_{\widetilde{2j}}.
\end{equation}
These modes transform into each other under the symmetry,
\begin{equation}
    \mathcal S \eta_L \mathcal S^{-1}=i\mathcal P \eta_R, \qquad \mathcal S \eta_R \mathcal S^{-1}=i\mathcal P \eta_L.
\end{equation}

The analysis above implies a fourfold degeneracy of all energy eigenstates in the thermodynamic limit\footnote{In a finite chain, the fourfold degeneracy divides into two degenerate doublets split by a gap which is exponentially small in the system size.} in the $J<h$ regime. By starting from an eigenstate with a sharp quantum number of the $\mathcal{S}$ symmetry (for example, $+1$), one can generate three additional orthogonal degenerate eigenstates by applying the operators $\eta$, $\eta'$ and $\eta' \eta\,$ to this state . As a result, the fourfold manifold contains states with all possible eigenvalues of the $\mathcal{S}$ symmetry, namely $+1, i, -1, -i$. The symmetry $\mathcal{S}$ enforces corresponding selection rules for allowed transitions within this manifold.

\subsubsection{Robustness of the Majorana strong zero modes} \label{sec:robust_zero}
\begin{table}
    \centering
    \ra{1.1}
    \begin{tabular}{crccrccrc}
    \toprule
     & & \multicolumn{2}{c}{$\mathcal H^+$} & & \multicolumn{2}{c}{$\mathcal H^-$} &  &\\
     \cmidrule{3-4} \cmidrule{6-7} 
         & & $\eta_1$ & $\eta$ &  & $\eta'$ & $\eta''$ & & degeneracy \\
    \midrule
         \multirow{ 2}{*}{$J>h$} & & exact & exact & & & & & \multirow{ 2}{*}{2}\\
         & & edge & interface & & & & & \\    
    \midrule
         \multirow{ 2}{*}{$J<h$} & & exact & exact & & approx. & approx. & & \multirow{ 2}{*}{4}\\
         & & edge & edges & & interface & edges & & \\
         \bottomrule
    \end{tabular}
    \caption{Strong zero modes in the open chain: For $J<h$ the modes $\eta$ and $\eta''$ can be combined to form $\eta_L$ and $\eta_R$, localized on the left and right edges respectively, see Eqs. \eqref{eq:eta_L} and \eqref{eq:eta_R}.}
    \label{tab:zm}
\end{table}

The properties of the strong zero modes in the open chain are summarized in Tab.~\ref{tab:zm}. 
We will now discuss the fate of these modes in the presence of small perturbations that either preserve or explicitly break the $\mathcal S$ symmetry. We assume the perturbations are small such that they do not close the bulk gap.
    
We first emphasize again that the strong zero mode $\eta_1$ remains exact for any local $\mathcal S$-invariant perturbation because, due to the reflection property $a\rightarrow\widetilde{a}$ of $\mathcal{S}\,,$ any term that couples $\eta_1$ with modes adjacent to the left end of the chain would be transformed into a non-local term that couples $\eta_1$ with the opposite end of the chain. Therefore, $\eta_1$ cannot appear in any local $\mathcal S$-invariant perturbation. For $J>h$ this is enough to ensure the robustness of the exact twofold degeneracy under such perturbations. We further know that for $J<h$, the modes $\eta_L$, $\eta_R$ and $\eta'$ are localized at the left edge, the right edge, and the interface respectively, so they cannot be coupled by any local perturbation in the thermodynamic limit. Hence, the fourfold degeneracy is also robust. Note that this is true both for quadratic and non-quadratic terms as long as they are $\mathcal S$-invariant, and does not depend on the decoupling of the $\xi^+$ and $\xi^-$ chains in the Hamiltonian.
In fact, while the property $\mathcal H =\mathcal H^+ +\mathcal H^-$ holds for generic $\mathcal S$-invariant quadratic Hamiltonians, interactions can couple the $\xi^+$ and $\xi^-$ chains.

If the $\mathcal S$ symmetry is explicitly broken, $\eta_1$ may be coupled to the rest of the chain. In the $J>h$ case, since $\eta_1$ is localized on the left edge while $\eta$ is localized at the interface, the twofold degeneracy is robust even to (local) perturbations that break $\mathcal S$. This degeneracy is not exact but holds up to exponentially small corrections. In the $J<h$ regime, the modes $\eta_1$ and $\eta_L\,,$ both localized on the left edge of the chain, can be coupled and gapped out by an $\mathcal S$-breaking perturbation, which lifts the degeneracy from fourfold to twofold in the thermodynamic limit. If, however, the perturbation does not contain $\eta_1$, the latter remains an exact strong zero mode and the discussion above valid for $\mathcal S$-invariant perturbations applies here as well, i.e., the degeneracy is not lifted.

While the dual-reflection interface studied in this paper is an integrable model that becomes quadratic in the fermion formulation, we can easily construct
different non-integrable deformations guided by the $\mathbb{Z}_4$ symmetry. For example, inspired by \cite{o2018lattice}, one can add a non-integrable term  $\sum_j Z_{j-1} X_j X_{j+1}$, which is invariant under the $\mathbb{Z}_4$ symmetry, to the Hamiltonian. In the fermion formulation, it is quartic in Majorana fermions.

\subsection{Closed chain}
\begin{figure}
    \centering
    \begin{tikzpicture}[scale=0.75]
    	\tikzstyle{every node}=[font=\fontsize{14}{14}\selectfont]
    	\draw[white] (-8.5,1) rectangle (9.7,-1); 
    	\begin{scope}[shift={(0,0)}]
    		\foreach \x in { -6,-2,2}{
    			\draw[orange, line width=3pt] (\x,0.7) -- (\x+2,0.7);	
    		}
    		
    		\foreach \x in { -4,0,4}{
    			\draw[green, line width=3pt] (\x,0.7) -- (\x+2,0.7);	
    		}
    		\draw[anthracite!70!black, line width=3pt] (-8,0.7) -- (-6,0.7);
    		\foreach \x in {-6,-4,...,6}
    		{        
    			\filldraw[fill=white] (\x,0.7) circle (0.2);
    			\draw[line width=1.2pt] (\x,0.55)--(\x,0.85);
    			\draw[line width=1.2pt] (\x-0.15,0.7)--(\x+0.15,0.7);
    		}
    		\filldraw[fill=white] (-8,0.7) circle (0.2);
    		\draw[line width=1.2pt] (-8,0.55)--(-8,0.85);
    		\draw[line width=1.2pt] (-7.85,0.7)--(-8.15,0.7);
    		
    		\draw (-1,0.7) node[fill=white]{...};
    		\draw (1,0.7) node[fill=white]{...};
    		\foreach \x in { -8,-6,-4,-2,2,4,6}{
    			\draw[<->, thick, blue] (\x,0.4)--(-\x,-1.7);
    		}
    		\begin{scope}[shift={(0,-1.5)}]
    			\foreach \x in { -6,-2,2}{
    				\draw[orange, line width=3pt] (\x,-0.7) -- (\x+2,-0.7);	
    			}
    			\foreach \x in { -4,0,4}{
    				\draw[green, line width=3pt] (\x,-0.7) -- (\x+2,-0.7);	
    			}
    			\draw[anthracite, line width=3pt] (6,-0.7) -- (8,-0.7);
    			
    			\draw (-1,-0.7) node[fill=white]{...};
    			\draw (1,-0.7) node[fill=white]{...};
    			
    			\foreach \x in {-6,-4,...,6}
    			{        
    				\filldraw[fill=white] (\x,-0.7) circle (0.2);
    				\draw[line width=1.2pt] (\x-0.15,-0.7)--(\x+0.15,-0.7);
    			}
    			\filldraw[fill=white] (8,-0.7) circle (0.2);
    			\draw[line width=1.2pt] (7.85,-0.7)--(8.15,-0.7);
    		\end{scope}
    		
    		\begin{scope}[shift={(0,-3)}]
    		\draw (-8,0) node[font=\fontsize{11}{11}]{$1$};
    		\draw (-6,0) node[font=\fontsize{11}{11}]{$2$};
    		\draw (6,0) node[font=\fontsize{11}{11}]{$N$};
    		\draw (8,0) node[font=\fontsize{11}{11}]{$N$$+$$1$};
    		\draw (0,0) node[font=\fontsize{11}{11}]{$a$};
    		\end{scope}
    	\end{scope}
    	\begin{scope}[shift={(7.5,1.2)}]
    		\draw[] (-0.2,0.4) rectangle (1.9,-2.2);
    		\draw[orange, line width=3pt] (0,0) -- (1,0);
    		\draw[] (1,0) node[anchor=west, font=\fontsize{11}{11}]{$J$};
    		\draw[green, line width=3pt] (0,-0.6) -- (1,-0.6);
    		\draw[] (1,-0.6) node[anchor=west, font=\fontsize{11}{11}]{$h$};
    		\draw[anthracite, line width=3pt] (0,-1.2) -- (1,-1.2);
    		\draw[] (1,-1.2) node[anchor=west, font=\fontsize{11}{11}]{$J_0$};
            \draw[anthracite!70!black, line width=3pt] (0,-1.8) -- (1,-1.8);
    		\draw[] (1,-1.8) node[anchor=west, font=\fontsize{11}{11}]{$J_0^\prime$};
    	\end{scope}	
    \end{tikzpicture}
    \caption{The two decoupled Majorana chains corresponding to the closed chain Hamiltonian \eqref{H_Majorana_closed}: The link between site 1 and site 2 of the $\xi^+$ chain represents the second (antiperiodic) interface with the coupling $J'_0$. The mapping \eqref{eq:map_close} indicated with blue arrows exchanges {$\xi_a^p$ and $\xi_{N+2-a}^{-p}$ Majoranas} such that the couplings change roles as $J\leftrightarrow h,\, J_0\leftrightarrow J_0^\prime\,.$}
    \label{fig:two_Majorana_closed}
\end{figure}
Next, we will consider a closed chain symmetric under the $\mathcal{S}$ transformation. Compared with the open chain discussed above, it has a second, (antiperiodic) interface, see Eq.~\eqref{H_Majorana_closed}. We will show that both interfaces host a localized strong zero mode.

First, notice that we can map the Hamiltonian $\mathcal{H}^\circ \equiv \mathcal{H}^\circ(J,h,J_0,J_0^\prime)$ onto a Hamiltonian with exchanged coupling strengths $\mathcal{H}^\circ(J\leftrightarrow h,\,J_0\leftrightarrow J_0^\prime)$ by the transformation
\begin{align}
    \eta_{a} \rightarrow \begin{cases}
    \eta_{N+a} & a\leq N\\
    -\eta_{a-N} & a > N
    \end{cases}
\end{align}
for $a=1,\dots,\,2N\,$,
which (up to a sign) corresponds to a translation of the Majorana operators by $N$ sites modulo $2N\,.$ 
It acts on $\xi_a^\pm$ as
\begin{align} \label{eq:map_close}
    \xi_a^p \rightarrow p{\,(-1)^{a+1}\,}\xi_{N+2-a}^{-p}
\end{align}
with $p=\pm \,$,
and thereby interchanges Majorana fermions as illustrated in Fig.~\ref{fig:two_Majorana_closed}. The strong zero modes of $\mathcal{H}^\circ(J\leftrightarrow h,\,J_0\leftrightarrow J_0^\prime)$ follow directly from those derived for $J>h$ using this mapping and exchanging $J\leftrightarrow h,\,J_0\leftrightarrow J_0^\prime$ in the given expressions.

{Without loss of generality, we will now focus on the regime $J>h\,,$ where the $\xi^+$ chain in Fig.~\ref{fig:two_Majorana_closed} is in the topological phase and thus hosts two Majorana strong zero modes.}
In the closed geometry, the interface-localized operator $\eta$ defined in Eq.~\eqref{eq:eta} is still a strong zero mode, but it commutes with $\mathcal{H}^\circ$ only up to an exponentially small error which vanishes in the thermodynamic limit.\footnote{ Here, we do not have the exact cancellation observed for the open chain at order $\pqty{\frac{h}{J}}^{n-1}$ since
$\bqty{\mathcal{H}^\circ,\, \xi_{2}^{+}} =  \bqty{\mathcal{H},\, \xi_{2}^{+}} + 2{\sqrt{2}} iJ_0^\prime\,\eta_1.$
It is the fact that the Ising coupling has different signs at both interfaces which hinders us from adding another term to get rid of the finite-size correction to the commutator.}

Now, we will construct the operator for the strong zero mode localized around the antiperiodic interface. We start from the ansatz
\begin{align}
    \eta^\prime = \frac{1}{\mathcal{N}^\prime}\bqty{\frac{1}{\sqrt{2}}\,{\xi_1^+} + \frac{J_0^\prime}{J} \, {\xi_3^+} + \mathcal{O}({\scriptstyle\frac{h}{J}})}\, .
\end{align}
By completing the iterative construction, we find the following operator, which commutes with the closed-chain Hamiltonian \eqref{H_Majorana_closed} up to a term exponentially small in system size,
\begin{align}
    \eta^\prime &= \frac{1}{\mathcal{N}^\prime}\bqty{ \frac{1}{\sqrt{2}} \, {\xi_1^+} + \frac{ J_0^\prime}{J}\sum_{j=1}^{n-1} \pqty{\frac{h}{J}}^{j-1}\,{\xi_{2j+1}^+} }\,,
\end{align}
with the normalization
\begin{align}
    {\mathcal{N}^\prime}^2 = \frac{1}{2} + {J_0^\prime}^2\,\frac{ 1-\pqty{\frac{h}{J}}^{N-2} }{ J^2-{h^2} }\,.
\end{align}
This Majorana mode converges to the local $\eta_1$ zero mode found in the open chain if the antiperiodic interface coupling is vanishingly small, $J_0^\prime \ll J\,.$
Under conjugation with the $\mathcal{S}$ symmetry, $\eta$ and $\eta^\prime$ transform in the same way, explicitly, $\mathcal{S}\,\eta \,\mathcal{S}^{-1} =  i\mathcal{P}\,\eta $ and $\mathcal{S}\,\eta^{\prime} \,\mathcal{S}^{-1} =  i\mathcal{P}\,\eta^{\prime} \,.$ We can combine them into a complex fermionic mode that creates exponentially close pairs of energy eigenstates throughout the spectrum.
This remains true even when the $\mathbb{Z}_4$ symmetry $\mathcal{S}$ is broken to the fermion parity $\mathcal{P}$ subgroup.

\subsection{Interface-localized strong zero modes in the spin chain}
After the inverse Jordan-Wigner transformation, the interface-localized zero modes $\eta$ in Eq.~(\ref{eq:eta}) and $\eta'$ in Eq.~\eqref{etapr} become non-local in terms of spin degrees of freedom. In the spin formulation of Sec. \ref{sec:spin}, they do therefore not form interface-localized strong zero modes.
But can the Ising spin chain host such interface-localized modes? This question has recently been studied in depth in \cite{olund2023boundary}, where the authors investigated interfaces in Ising and Kitaev chains.\footnote{
In \cite{olund2023boundary}, the Ising and transverse field couplings are $J_1,\,h_1$ and $J_2,\,h_2$ on the left and right, respectively.
In the parameter subspace $J_1 = h_2,\, h_1 = J_2\,$, the model studied in \cite{olund2023boundary} corresponds to our Kramers-Wannier dual interface tuned to $J_0 = J\,.$} While the interface between Kitaev chains hosts a localized strong zero mode if and only if there is a phase boundary, the existence criterion for the localized zero mode is more complex in the Ising model. We observe here that the perturbative iterative construction of \cite{olund2023boundary} suffers from a resonance divergence when fine-tuned to the dual-reflection interface. This suggests the absence of an exponentially-localized interface strong zero mode in the spin formulation.

\subsection{Summary}
For the open chain, the $\mathcal{S}$ symmetry and locality together imply a pair of exact Majorana strong zero modes, which ensure that all energy levels have an exact twofold degeneracy even for finite chains. This degeneracy remains robust under local $\mathbb{Z}_4$ symmetry-preserving interactions. In the thermodynamic limit,
we have demonstrated that the number of Majorana strong zero modes in the open chain and the related degeneracy of energy eigenstates changes as we cross the critical point $J=h$. While in the $J>h$ regime there is a twofold degeneracy, in the $J<h$ regime we identified four Majorana strong zero modes, which jointly entail a fourfold degeneracy of eigenstates. The symmetry $\mathcal{S}$ readily distinguishes the four orthogonal basis states and forbids transitions between them. The difference in ground state degeneracies between the two regimes also persists in the presence of local symmetry-preserving interactions. Note, however, that the ground state degeneracy for $J<h$ is lifted to twofold by locally hybridizing the two Majorana zero modes $\eta_1$ and $\eta_L$, which necessarily breaks the $\mathcal{S}$ symmetry.

With closed boundary conditions, by contrast, we have shown how only a single pair of Majorana strong zero modes can be constructed for arbitrary (non-critical) choices of coupling constants. Both modes live in the same $\mathcal{S}$ symmetry sector and possess an exponentially small energy suppressed by the system size, which gives rise to a doubly degenerate energy spectrum in the thermodynamic limit. Breaking the $\mathcal{S}$ symmetry does not change this qualitatively.

\section{Realization in digital quantum simulators}\label{sec:quantum circuit}

\begin{figure}[t]
    \centering
    \hspace*{-8pt}
    \begin{tikzpicture}[scale=0.77]
    	\tikzstyle{every node}=[font=\fontsize{14}{14}\selectfont]
    	\foreach \x in {-6,-4,...,8}
    	{        
    		\filldraw (\x,0) circle (0.1);
    		\filldraw (\x,0) -- (\x,5.4);
    	}
    	\foreach \x in {-6, -2}
    	{        
    		\filldraw[orange, fill=white, line width=3pt] (\x-0.2, 3.7) rectangle (\x+2.2, 4.7);
    		\draw (\x+1,4.2) node[]{$R_{XX}(\theta)$};
    	}
    	\foreach \x in {-4}
    	{        
    		\filldraw[orange, fill=white, line width=3pt] (\x-0.2, 2.2) rectangle (\x+2.2, 3.2);
    		\draw (\x+1,2.7) node[]{$R_{XX}(\theta)$};
    	}
    	\foreach \x in {0}
    	{        
    		\filldraw[anthracite, fill=white, line width=3pt] (\x-0.2, 2.2) rectangle (\x+2.2, 3.2);
    		\draw (\x+1,2.7) node[]{$U_{J_0}$};
    	}
    	\foreach \x in {4}
    	{        
    		\filldraw[green, fill=white, line width=3pt] (\x-0.2, 2.2) rectangle (\x+2.2, 3.2);
    		\draw (\x+1,2.7) node[]{$R_{XX}(\varphi)$};
    	}
    	\foreach \x in {2, 6}
    	{        
    		\filldraw[green, fill=white, line width=3pt] (\x-0.2, 0.7) rectangle (\x+2.2, 1.7);
    		\draw (\x+1,1.2) node[]{$R_{XX}(\varphi)$};
    	}
    	\foreach \x in {4, 6, 8}
    	{        
    		\draw[orange, fill=white, line width=3pt] (\x-0.9, 3.7) rectangle (\x+0.9, 4.7);
    		\draw (\x,4.2) node[]{$R_{Z}(\theta)$};
    	}
    	\foreach \x in {-6, -4, -2}
    	{        
    		\draw[green, fill=white, line width=3pt] (\x-0.9, 0.7) rectangle (\x+0.9, 1.7);
    		\draw (\x,1.2) node[]{$R_{Z}(\varphi)$};
    	}
    	\foreach \x in {-6, 8}
    	{   
    		\filldraw[white, fill=white] (\x-0.4, 2.2) rectangle (\x+0.4, 3.2);     
    		\draw (\x,2.7) node[]{...};
    	}
    
    	\draw [decorate,decoration={brace,amplitude=5pt}, line width=1.5pt]
    	(-7.2,0.4) -- (-7.2,5.1) node[midway,xshift=-15pt]{$ U $};
    	
    
    	\begin{scope}[shift={(9.5,5)}]
    		\draw[] (-0.15,0.4) rectangle (1.75,-1.6);
    		\draw[orange, line width=3pt] (0,0) -- (1,0);
    		\draw[] (1,0) node[anchor=west, font=\fontsize{11}{11}]{$J$};
    		\draw[green, line width=3pt] (0,-0.6) -- (1,-0.6);
    		\draw[] (1,-0.6) node[anchor=west, font=\fontsize{11}{11}]{$h$};
    		\draw[anthracite, line width=3pt] (0,-1.2) -- (1,-1.2);
    		\draw[] (1,-1.2) node[anchor=west, font=\fontsize{11}{11}]{$J_0$};
    	\end{scope}	
    \end{tikzpicture}
    \caption{Quantum circuit for $S$-preserving time evolution: We apply three layers of one- and two-qubit gates which implement together the time evolution operator $U$ corresponding to a Trotter time step. The gates $R_Z,\, R_{XX}$ and $U_{J_0}$ are specified in Eqs.~\eqref{eq:RZ}, \eqref{eq:RXX} and \eqref{eq:UJ0} respectively, and their phases are $\theta =J\tau$, $\varphi=h\tau$ and $\alpha=J_0\tau$, in which $\tau$ gives the time step. The colors of the blocks are chosen according to the coupling strength appearing in the respective gates. For each color, applying the complete set of corresponding gates preserves the $S$ symmetry.}
    \label{fig:quantum_circuit}
\end{figure}

The presence of zero modes can be probed experimentally in quantum simulators, manifesting in a slow decay of specific expectation values, as has been demonstrated on both analog and digital platforms \cite{Google2022, jin_observation_2025}. The simulated models correspond to the spin analogs of fermionic symmetry protected topological phases with Majorana edge modes. Even though these spin models are genuinely robust only to symmetric noise, they were also shown to be long-lived against certain symmetry-breaking noise thanks to the mechanism of prethermalization \cite{fendley_prethermal,Google2022}. We may therefore expect a similar prethermal robustness for an experimental implementation of the dual-reflection interface.

In digital quantum devices, continuous Hamiltonian evolution can be approximated via a Trotter decomposition in the limit of small time steps. However, this approach requires a large number of gates, making it more susceptible to errors. A more practical alternative is to use a moderately large time step while preserving the key properties of the model, such as the invariance under symmetries. Quantum circuits of this form can still support Majorana strong zero modes, which commute with the unitary operator representing the evolution over one cycle. Away from the limit of a small Trotter step, these models can also exhibit more exotic edge modes, including $\pi$-modes, which anticommute with the evolution operator \cite{Thakurathi2013,Yates2019}.

In this section, we demonstrate how to construct discrete-time realizations of our model, in the spin and fermion formulations, that preserve the $\mathbb{Z}_4$ symmetry, and we show that they can support strong zero modes. 

Specifically, within the spin formulation we implement a quantum circuit corresponding to the following time evolution operator (see Fig.~\ref{fig:quantum_circuit}):

\begin{equation}  \label{eq:U_dig}
    U= U_h \, U_{J_0} \, U_J \,
\end{equation}
with
\begin{align}
    U_J &= \prod_{j=1}^{n-1} R_{XX}(\theta)_{j,j+1} \prod_{j=n+2}^N R_Z(\theta)_j\,,\\
    U_h &= \prod_{j=n+1}^{N-1} R_{XX}(\varphi)_{j,j+1} \prod_{j=2}^{n} R_Z(\varphi)_j\, ,\\ \label{eq:UJ0}
    U_{J_0} &= \exp\left(-i\,\frac{\alpha}{2}\,(X_{n} X_{n+1}+Z_{n+1})\right) \,,
\end{align}
where the one-qubit gate $R_Z$ and the two-qubit gate $R_{XX}$ are defined as
\begin{align}\label{eq:RXX}
    R_{XX}(\theta)_{j,j+1} &= \exp\left(-i\,\frac{\theta}{2}\, X_j X_{j+1}\right)\,,\\ \label{eq:RZ}
    R_{Z}(\theta)_{j} &= \exp\left(-i\,\frac{\theta}{2}\, Z_j\right).
\end{align}
As shown in Fig.~\ref{fig:quantum_circuit}, by treating $R_{XX}, R_Z$ and $U_{J_0}$ as native one- and two-qubit gates, the evolution over a single cycle can be implemented using a quantum circuit of depth three. A more convenient set of quantum gates would be obtained from swapping $X$ and $Z\,.$ With the new notation, both $R_{XX}$ and $U_{J_0}$ could be decomposed into the product of a single-qubit gate and a controlled unitary \cite{kim2023evidence}.
However, for consistency, we will maintain the notation used in this paper.

Continuous time-Hamiltonian evolution with the Hamiltonian (\ref{H_int}) is recovered for $\theta =J\tau$, $\varphi=h\tau$, $\alpha=J_0\tau$ in the limit $\tau\rightarrow 0$. Away from this limit, the circuit $U$ retains many of the key features of interest, particularly its invariance under $S$ which is immediate to check, as each of the terms $U_J$, $U_h$, and $U_{J_0}$ is clearly invariant. 

Now, we construct the Majorana strong zero modes in the discrete-time Floquet model. Following the convention in \cite{Google2022}, we set $\tau=\pi\,$. Using the non-local Majoranas $\xi^\pm$, defined in Eq.~\eqref{eq:xi}, we first express the unitary circuit \eqref{eq:U_dig} as a product of two commuting operators,
\begin{equation}
	\mathcal{U} = \mathcal{U}_+\,\mathcal{U}_-\,,\qquad \bqty{\mathcal{U}_+,\,\mathcal{U}_-}=0\,,
\end{equation}
with the definitions
\begin{align}
    \mathcal{U}_{+} &=  \prod_{j=2}^{n} \exp{-\frac{\pi h}{2}\, \xi^+_{2j-1}\xi^+_{2j}} \prod_{j=1}^{n-1} \exp{-\frac{\pi J}{2}\, \xi^+_{2j}\xi^+_{2j+1}} \,,\\
	\mathcal{U}_- &= \prod_{j=2}^{n} \exp{-\frac{\pi h}{2}\, \xi^-_{2j-1}\xi^-_{2j}}\, \exp\Bqty{-\frac{\pi J_0}{\sqrt{2}}\,\xi^-_{N}\xi^-_{N+1}} \prod_{j=1}^{n-1} \exp{-\frac{\pi J}{2}\, \xi^-_{2j}\xi^-_{2j+1}} \,.
\end{align}
We define a Floquet strong zero mode to be a Majorana operator $\Psi$,
\begin{equation}
	\Psi = \sum_{j=1}^{N} \psi_{2j-1}\,\eta_{2j-1} + \psi_{2j}\, \eta_{2j} \,,
\end{equation}
which is odd under fermion parity $\mathcal{P}$ and invariant under Floquet evolution, i.e it commutes with the evolution operator $\mathcal{U}$,
\begin{equation} 
	\mathcal{U}^{-1}\,\Psi\,\mathcal{U} = \Psi\,.
\end{equation}
Since it is specialized to the case in which the leftmost coupling vanishes, $h_1 = 0\,,$ we find that the operator $\mathcal{U}_{+}$ exactly commutes with the Majorana mode
\begin{equation}\label{eq:interface FMZM}
\begin{split}
	\Psi =\, \frac{1}{\mathcal{N}}\,&\Bigg[ \sum_{j=2}^{n} {\lambda_0}^{n-j} \pqty{-\sin(\frac{\pi h}{2})\, \xi^+_{2j-1} + \cos(\frac{\pi h}{2})\, \xi^+_{2j}} + {\lambda_0}^{n-1} \cos(\frac{\pi h}{2}) \, \xi^+_{2}  \Bigg]\,,
\end{split}
\end{equation}
with
\begin{align}\label{eq:lambda0}
    \lambda_0 &= \frac{\tan(\pi h/2)}{\tan(\pi J/2)}\,,\\
    \mathcal{N}^2 &= 2 \cos(\frac{\pi J}{2})^2 \frac{1-{\lambda_0}^2}{1 + \cos(\pi J) - 2\cos(\pi h/2)^2 \,{\lambda_0}^N}\,.
\end{align}
Since $\mathcal{U}_-$ commutes with $\Psi$ trivially, we conclude that this Majorana operator is an exact strong zero mode. Remembering that the $\xi^+$ are defined on the folded + chain, which contains the interface as link between the Majorana sites $N$ and $N+1$, we see that $\Psi$ is localized around the dual-reflection interface if $\lambda_0 < 1\,,$ and on the edges if $\lambda_0 > 1\,.$ It is paired with the exact Floquet Majorana strong zero mode
$\Psi_1 = \xi^+_1$\,,
which does not appear in the evolution operator.  Together, these modes imply a long-lived fermionic excitation that commutes with the Floquet evolution exactly, even on chains of finite length.\footnote{
Together with the exact zero modes, the $\xi^+$ chain supports (approximate) Floquet Majorana strong $\pi$ modes for $|\tan(\pi h/2)\tan(\pi J/2)|>1$ \cite{Thakurathi2013,Yates2019}. The defining property of these modes is that they are spatially localized, odd under fermion parity $\mathcal P$ and that they change sign under the Floquet evolution,
$\mathcal U^{-1} \,\Psi_\pi\, \mathcal U=-\Psi_\pi$,
up to exponentially small corrections in the system size. Similarly, the $\xi^-$ chain supports approximate 0 and $\pi$ modes for $|\tan(\pi J/2)/\tan(\pi h/2)|<1$ and $|\tan(\pi h/2)\tan(\pi J/2)|>1$, respectively.
}

Since the limit $\tau \to 0$ corresponds to infinitesimal Hamiltonian time evolution, Floquet zero modes can be deformed into exact zero modes of the Hamiltonian by taking the limit $\tau\to 0\,.$ It is easy to check that in this limit one recovers from Eq. \eqref{eq:interface FMZM} the exact strong zero mode \eqref{eq:eta} on the $+$ chain.

In summary, we constructed a quantum circuit enjoying the $\mathbb{Z}_4$ dual-reflection symmetry. The symmetry protects Floquet Majorana strong zero modes that are exact even on chains of a finite length. 
Therefore, the discrete-time Floquet framework provides a practical pathway towards the realization of such modes in current digital quantum simulators.


\section{Conclusions and outlook} \label{sec:outlook}
In this paper, we investigated an interface between an Ising paramagnetic and ferromagnetic region that is symmetric under a composition of the Kramers-Wannier transformation and spatial reflection. This gives rise to a dual-reflection symmetry that is best understood in the fermion language. After Jordan-Wigner transformation, the symmetry manifests as parity-dependent spatial reflection symmetry across a site of the Majorana chain. Its symmetry operator obeys a $\mathbb{Z}_4$ algebra and squares to the $\mathbb{Z}_2$ fermion parity. In the spin language, the open chain with dual-reflection interface enjoys a $\mathbb{Z}_4$ symmetry, while introducing end-to-end coupling in the form of a second dual-reflection interface makes the symmetry non-invertible. In contrast to the interface corresponding to the topological Kramers-Wannier non-invertible defect, the construction of the dual-reflection interface does not assume criticality - providing insights into non-invertible symmetries at interfaces between gapped phases. 

The dual-reflection symmetry protects strong Majorana zero modes localized on the interface and edges of the system. In particular, we uncovered exactness of certain zero modes. The strong zero modes moreover lead to a degeneracy pattern that is tunable between two- and fourfold degeneracy and spans the whole manifold of energy eigenstates. While the dual-reflection interface studied in this paper is an integrable model,
we showed that the strong zero modes cannot be lifted by local, symmetry-preserving interactions. We even made the first steps towards the realization of the Ising dual-reflection interface in digital quantum simulators by constructing quantum circuits in the spin and fermion formulations. Also for the corresponding Floquet evolution, we explicitly derived exact strong Majorana zero modes in the discrete setup, protected by the symmetry.

Due to these zero modes, one possible application of the dual-reflection symmetric chain is as a platform for qubit engineering. The exact strong Majorana zero modes, as well as their Floquet counterparts, could be used to build long-lived qubits that are fractionalized between the edges and the interface, even on chains of modest lengths. These qubits would also be robust to symmetry-preserving perturbations. The exactness of the zero modes could lower hardware overheads and enable a compact qubit implementation, as it removes the usual need for long chains to suppress Majorana splitting. 
In practice, this however requires fine-tuning imposed by the dual-reflection symmetry.
In the \(J<h\) regime, we identified four Majorana strong zero modes. In parity-preserving quantum error correction codes, four such modes are needed to encode one logical qubit \cite{bravyi_majorana_2010}. These Majorana fermion codes usually exploit the non-Abelian statistics of Majoranas in two spatial dimensions to create qubits that are robust to local errors. Recently, it has been shown that this braiding can also be mimicked in one-dimensional systems by combining different Majorana eigenstates of the Floquet evolution (for example zero- and $\pi$ modes) \cite{bomantara_simulation_2018,liu_floquet_2013,bomantara_quantum_2018,bauer_topologically_2019}. An increased number of available Majorana modes can be another advantage of the Floquet setting. It would be interesting to investigate whether short, dual-reflection symmetric chains could be used as physical building blocks of Majorana quantum error correction codes.


Several present-day analog and digital quantum simulators offer local control over the individual couplings and can realize tunable Ising-type Hamiltonians. Therefore, we foresee possible implementations in various setups such as superconducting qubits \cite{Google2022}, trapped ions \cite{Monroe2021,de2024observation} and neutral atoms \cite{Steinert2023}. Our symmetry-preserving circuit implementation is of depth three, and therefore suitable for Noisy Intermediate-Scale Quantum devices.

Prethermalization has been shown to protect boundary strong zero modes from scattering with thermally excited electronic quasiparticles \cite{fendley_prethermal, jin_observation_2025}. In the dual-reflection symmetric chain, due to its symmetry, domain wall and spin flip bulk excitations in the ferromagnetic, respectively paramagnetic part of the chain have the same energy at given momentum. This raises the following questions: How can we understand the physical mechanisms governing scattering in a dual-reflection symmetric chain? Which role does the dual-reflection symmetry play for scattering between bulk excitations and edge- and interface modes?
Answering these questions will not only clarify the long time and finite temperature behavior of our model but also yield conceptual insights into the physics at one-dimensional phase boundaries. A step into this direction was made in \cite{ueda_perfect_2025}, where an interface between Kramers-Wannier dual regions of a spin chain was shown to perfectly transmit low-energy excitations from the ferromagnetic into the paramagnetic region. The latter interface becomes the topological defect of the Kramers-Wannier symmetry at criticality. It is therefore not equivalent to our symmetric interface, and it will be interesting to investigate how this difference manifests itself. By virtue of the dual-reflection symmetry, we expect to be able to obtain exact analytic results, making the dual-reflection interface an attractive test bed for studies of quantum dynamics.

Generalized symmetries are known to be closely related to corresponding topological defects\cite{affleck1997boundary, PhysRevLett.93.070601, grimm2002spectrum, frohlich2007duality, aasen2016topological, aasen2020topological, PhysRevB.94.115125, seiberg2024majorana, seiberg2024non, carqueville2023topological}. From this perspective, it is instructive to search for topological defects associated with the dual-reflection symmetry and to determine their fusion rules. To this end, we intend to use the folding method, which has proven instrumental in the study of conformal interfaces in the Ising conformal field theory \cite{affleck1997boundary}.

\section*{Acknowledgements}
We are grateful to Kristian Tyn Kai Chung, J\"urgen Fuchs, Hosho Katsura, Ho Tat Lam, Nandago\-pal Manoj, Per Moosavi, Masaki Oshikawa, Sal Pace, Abhinav Prem, Apoorv Tiwari, Konstantin Zarembo for discussions and comments.


\paragraph{Funding information}
S.M. and J.G. are supported by Vetenskapsr{\aa}det (2021-03685), Carl Tryggers Stiftelse (CTS 24:3607), Wenner-Gren Stiftelserna (UPD2024-0111) and STINT. M.B and S.M acknowledge support provided by Nordita. F.M.S.~acknowledges support provided by the U.S. Department of Energy (DOE) QuantISED program through the theory consortium ``Intersections of QIS and Theoretical Particle Physics'' at Fermilab, and by Amazon Web Services, AWS Quantum Program.

\begin{appendix}
\numberwithin{equation}{section}

\section{Kramers-Wannier transformation} \label{appA}
In this appendix, we introduce aspects of the Kramers-Wannier transformation applied to the transverse field Ising chain which are mentioned in the main text.
The Kramers-Wannier transformation is a non-local map acting on a spin $1/2$ chain. We perform it by conjugating spin operators with the following unitary operator, see for example \cite{shao2023s, seiberg2024non} and references therein, on a chain with $N$ sites:\footnote{The index 
$j$ runs from $N$ to $2$ from left to right.}
\begin{equation}
    \op{U}{KW}=e^{-2\pi i N/8} \left(\prod_{j=N}^{2} \frac{1+i \op{Z}{j}}{\sqrt{2}} \frac{1+i \op{X}{j} \op{X}{j-1}}{\sqrt{2}} \right) \frac{1+i \op{Z}{1}}{\sqrt{2}}\,.
\end{equation}
Consider now its action on the local spin operators which appear in the Hamiltonian of the transverse field Ising chain. The leftmost transverse-field operator $Z_1$ and the end-to-end coupling operator $X_1 X_N$ are mapped onto non-local operators,
\begin{equation}
    U_{KW}\, \op{Z}{1}\, U_{KW}^{-1} = Q \op{X}{1} \op{X}{N}, \qquad U_{KW}\, \op{X}{1} \op{X}{N} U_{KW}^{-1} = Q \op{Z}{N}\,,
\end{equation}
where $Q=\prod_{j=1}^{N} \op{Z}{j}$ denotes the Ising symmetry operator. On the remaining operators, the Kramers-Wannier unitary $U_{KW}$ acts as
\begin{equation} \label{eq:KW_bulk}
    U_{KW}\, \op{Z}{j}\, U_{KW}^{-1} = \op{X}{j-1} \op{X}{j}, \qquad U_{KW}\, \op{X}{j-1} \op{X}{j}\, U_{KW}^{-1} = \op{Z}{j-1} \quad \text{for } j=2,3,\dots, N\,.
\end{equation}
These transformation properties show in particular that the Ising symmetry commutes with the Kramers-Wannier transformation, i.e., 
$\bqty{Q,\,U_{KW}} = 0\,$.
From Eq. \eqref{eq:KW_bulk} we observe that acting with $U_{KW}$ twice is, in the bulk, equivalent to a translation to the left by one lattice site.

On a periodic chain, the non-invertible operator that projects the Kramers-Wannier unitary on the $Q=+1$ parity sector,
\begin{equation} \label{D_ni}
    D=\op{U}{KW} \frac{1+Q}{2}\,,
\end{equation}
commutes with the transverse field Ising Hamiltonian tuned to the critical point,
\begin{equation}
    \op{H}{cr}=-\sum_{j=1}^{N} \op{X}{j}\op{X}{j+1}-\sum_{j=1}^{N} \op{Z}{j}.
\end{equation}
Thus, $D$ is the operator that implements the Kramers-Wannier non-invertible symmetry of the critical Ising model on a periodic chain. Note that while the unitary $\op{U}{KW}$ explicitly breaks translation symmetry, the non-invertible symmetry $D$ does not. Although this is not manifest from Eq.~\eqref{D_ni}, it becomes apparent from the matrix product operator expression \cite{seiberg2024non}.

For open boundary conditions, the Kramers-Wannier transformation is not a symmetry of the transverse field Ising model in any Ising parity sector.

\section{Kramers-Wannier and spatial reflection transformation in the\\ Majorana fermion formulation} \label{appB}
This section complements Sec.~\ref{sec:ferm_Z4} by providing details on the implementation of the Kramers-Wannier and spatial reflection transformations in the Majorana fermion formulation.

\subsection{Kramers-Wannier transformation}\label{appB-KW}
After the Jordan-Wigner transformation \eqref{eq:JW_Maj}, the unitary Kramers-Wannier operator \eqref{KW} and its inverse become
\begin{equation}
\begin{split}
	\mathcal{U}_{KW} &= e^{-2\pi i N/8}\, \frac{1+\op{\eta}{2N-1}\op{\eta}{2N}}{\sqrt{2}}\,\dots\, \frac{1+\op{\eta}{2}\op{\eta}{3}}{\sqrt{2}} \frac{1+\op{\eta}{1}\op{\eta}{2}}{\sqrt{2}}\, ,\\
	\mathcal{U}_{KW}^{-1} &= e^{2\pi i N/8}\, \frac{1-\op{\eta}{1}\op{\eta}{2}}{\sqrt{2}} \frac{1-\op{\eta}{2}\op{\eta}{3}}{\sqrt{2}}\,\dots\, \frac{1-\op{\eta}{2N-1}\op{\eta}{2N}}{\sqrt{2}} \,,
 \end{split}
\end{equation}
with individual factors acting as Majorana SWAP gates,
\begin{align}
	\frac{1+\op{\eta}{a}\op{\eta}{b}}{\sqrt{2}}\,\op{\eta}{c}\,\frac{1-\op{\eta}{a}\op{\eta}{b}}{\sqrt{2}} &= e^{\frac{\pi}{4}\op{\eta}{a}\op{\eta}{b}}\,\op{\eta}{c}\, e^{-\frac{\pi}{4}\op{\eta}{a}\op{\eta}{b}}=\nonumber\\
	& = -\,\delta_{a,c}\,\op{\eta}{b} + \delta_{b,c}\,\op{\eta}{a} + (1-\delta_{a,c})(1-\delta_{b,c}) \,\op{\eta}{c}\, ,
\end{align}
where we assume that $a\ne b\,.$
The unitary operator $\mathcal{U}_{KW}$ describes a Majorana sequential circuit, see for example \cite{PhysRevB.109.075116}. On individual Majorana operators, it acts as follows,
\begin{equation}
\begin{split}
        \mathcal{U}_{KW}\,\op{\eta}{1}\,\mathcal{U}_{KW}^\dagger &= - \op{\eta}{2N}\, , \\
	\mathcal{U}_{KW}\,\op{\eta}{a}\,\mathcal{U}_{KW}^\dagger &=  \op{\eta}{a-1}\,\qquad \text{for }  a=2,3,\dots, 2N \, .
\end{split}
\end{equation}
We can check that the action of the Majorana sequential circuit $ \mathcal{U}_{KW} $ in the fermion model agrees with that of the Kramers-Wannier unitary $ U_{KW} $ in the spin model. Using Eq.~\eqref{eq:KW_M}, we find the following correspondences in the bulk:
\begin{equation}
\begin{split}
	\mathcal{U}_{KW} \,\op{\eta}{2j-1}\op{\eta}{2j} \, \mathcal{U}_{KW}^\dagger = \op{\eta}{2j-2}\op{\eta}{2j-1} \quad &\leftrightarrow \quad U_{KW}\,\op{Z}{j}\,U_{KW}^\dagger = \op{X}{j-1}\op{X}{j} \,,\\
	\mathcal{U}_{KW} \,\op{\eta}{2j}\op{\eta}{2j+1} \, \mathcal{U}_{KW}^\dagger = \op{\eta}{2j-1}\op{\eta}{2j} \quad &\leftrightarrow \quad U_{KW}\,\op{X}{j}\op{X}{j+1}\,U_{KW}^\dagger = \op{Z}{j}\,,
 \end{split}
\end{equation}
where the right hand sides of the two equations in each line are related by the Jordan-Wigner transformation, as expected.
Note that for the exact strong zero mode at the left edge in spin and fermion formulation, we consistently find
\begin{align}
	\mathcal{U}_{KW}\,\op{\eta}{1}\,\mathcal{U}_{KW}^\dagger = - \op{\eta}{2N}\quad\leftrightarrow\quad U_{KW} \,\op{X}{1}\,U_{KW}^\dagger = i\,Q\,\op{X}{N} \,.
\end{align}


\subsection{Spatial reflection transformation}
We will now demonstrate that under the spatial reflection $\mathcal{R}$  swapping the sites $j\leftrightarrow N+1-j$, the Majorana fermions transform as
\begin{equation} \label{eq:eta_ref}
\begin{split}
	\mathcal{R}\, \op{\eta}{2j-1} \, \mathcal{R}^{-1} &=  i \mathcal{P} \, \op{\eta}{\widehat{2j-1}}\,,\\
	\mathcal{R}\,\op{\eta}{2j} \,\mathcal{R}^{-1} &= \ -i \mathcal{P} \, \op{\eta}{\widehat{2j}}\, ,
\end{split}
\end{equation}
where we introduced the notation $\widehat{a}=2N+1-a\,.$ Clearly, $\mathcal{R} \mathcal{P} \mathcal{R}^{-1}=\mathcal{P}$ and $\mathcal{R}^2 \op{\eta}{a} \mathcal{R}^{-2}=\op{\eta}{a}\,.$

The transformations \eqref{eq:eta_ref} are consistent with how spin operators $X_j$ and $Z_j$ should transform under spatial reflections. Indeed, using the Jordan-Wigner transformation
\begin{align}
    \mathcal{R}\, \underbrace{\left( -i \op{\eta}{2j-1}\op{\eta}{2j} \right)}_{JW(\op{Z}{j})} \,\mathcal{R}^{-1} &= -i \left( i \mathcal{P} \op{\eta}{\widehat{2j-1}}  \right) \left( -i \mathcal{P} \op{\eta}{\widehat{2j}} \right) =\nonumber\\ &= 	\underbrace{-i \op{\eta}{2(N+1-j)-1}\op{\eta}{2(N+1-j)}}_{JW(\op{Z}{N+1-j})}\, .
\end{align}
Moreover, with the previous result and the equality $(-1)^{\op{n}{j}} = -i \op{\eta}{2j-1}\op{\eta}{2j}\,,$ we derive
\begin{align}
	\mathcal{R} \, \underbrace{\prod_{k=1}^{j-1} (-1)^{\op{n}{k}} \, \op{\eta}{2j-1} }_{JW(\op{X}{j})} \,\mathcal{R}^{-1} &= \prod_{k=1}^{j-1} (-1)^{\op{n}{N+1-k}} \,i\mathcal{P}\, \op{\eta}{\widehat{2j-1}}=\nonumber\\
	&= i \prod_{k=1}^{N+1-j} (-1)^{\op{n}{k}}\,\op{\eta}{2(N+1-j)} =\nonumber\\
	&= i \prod_{k=1}^{N-j} (-1)^{\op{n}{k}} (-i\,\op{\eta}{2(N+1-j)-1}\op{\eta}{2(N+1-j)})\,\op{\eta}{2(N+1-j)}=\nonumber\\
	&= \underbrace{\prod_{k=1}^{N-j} (-1)^{\op{n}{k}}\,\op{\eta}{2(N+1-j)-1}}_{JW(\op{X}{N+1-j})}\,.
\end{align}
Note that this second condition on $\mathcal{R}$ makes the appearance of the parity operator in the action of $\mathcal{R}$ necessary and moreover fixes the phase. 

While it is not needed in the main text, we can obtain an explicit form of the reflection operator $\mathcal R$ as the Jordan-Wigner transformation of the reflection operator $R$ in spin language. For a chain of even length $N=2n\,,$ the latter is a product of $n$ two-qubit SWAP gates,
\begin{equation}
    R = \prod_{j=1}^n \frac{1}{2}\,\pqty{1 + X_j X_{N+1-j} + Y_j Y_{N+1-j} + Z_j Z_{N+1-j}}\,,
\end{equation}
and its Majorana formulation can be deduced from the Jordan-Wigner transformation of the spin $1/2$ Pauli operators to products of Majorana fermions,
\begin{equation}
\begin{split}
    Z_j &\rightarrow -i\,\eta_{2j-1}\eta_{2j} \,,\\
    X_j &\rightarrow \prod_{k=1}^{j-1} \pqty{-i\,\eta_{2k-1}\eta_{2k}} \, \eta_{2j-1} \,,\\
    Y_j &\rightarrow \prod_{k=1}^{j-1} \pqty{-i\,\eta_{2k-1}\eta_{2k}} \, \eta_{2j} \,.
\end{split}
\end{equation}

 \section{The dual-reflection interface with an alternative definition of the Kramers-Wannier transformation}\label{app:alternative_KW}
On a periodic chain, the square of the unitary Kramers-Wannier operator \eqref{KW} acts on local operators that commute with the Ising parity as a lattice translation to the left by one site, e.g. $Z_j\to Z_{j-1}$ for $j=2,\dots,N\,.$ Thus, the Kramers-Wannier transformation can be loosely described as a half-translation. This becomes manifest in the Majorana language, where acting once with the Kramers-Wannier unitary implements an elementary translation $\eta_a \to \eta_{a-1}$ on a doubled chain with $2N$ Majorana fermions that satisfy antiperiodic boundary conditions. The direction of the translations involved in both formulations arises from an arbitrary choice we made in the definition of the Kramers-Wannier operator. In  this appendix, we explore an implementation of the Kramers-Wannier transformation alternative to the one used in the main text.

Let us swap the order of the individual factors and define
\begin{equation}
    U_{KW}^\prime = e^{-{2\pi i N}/8}\, \left(\prod_{j=1}^{N-1} \frac{1+i \op{Z}{j}}{\sqrt{2}} \frac{1+i \op{X}{j} \op{X}{j+1}}{\sqrt{2}} \right) \frac{1+i \op{Z}{N}}{\sqrt{2}}\,.
\end{equation}
This unitary operator maps the following operators onto non-local ones,
\begin{equation}
    {U}_{KW}^\prime\, Z_N \,({U}_{KW}^\prime)^{-1} = Q\,X_N X_1 \,,\qquad {U}_{KW}^\prime\, X_N X_1 \,({U}_{KW}^\prime)^{-1} =  Q\, Z_1 \,,
\end{equation}
and acts on the other operators appearing in the Hamiltonian of the transverse field Ising chain as
\begin{equation}
    {U}_{KW}^\prime\, Z_j \,({U}_{KW}^\prime)^{-1} = X_j X_{j+1} \,,\qquad  {U}_{KW}^\prime\, X_j X_{j+1} \,({U}_{KW}^\prime)^{-1} = Z_{j+1}\,,
\end{equation}
for $j=1,2,\dots, N-1$ (compare with Eq.~\eqref{KW2} for our convention).
Consequently, its square equals a lattice translation to the right by one spin site. The Jordan-Wigner dual operator is, expressed in terms of Majorana fermions,
\begin{equation}
    \mathcal{U}_{KW}^\prime = e^{-{2\pi i N}/8}\,\prod_{a=1}^{2N-1} \frac{1+\eta_a\eta_{a+1}}{\sqrt{2}}\,,
\end{equation}
and we find the transformation
\begin{equation}
\begin{split}
    \mathcal{U}_{KW}^\prime\, \eta_a \, (\mathcal{U}_{KW}^\prime)^{-1} &= -\eta_{a+1} \quad \text{for } a = 1,\dots,2N-1\,,\\ \mathcal{U}_{KW}^\prime\, \eta_{2N} \, (\mathcal{U}_{KW}^\prime)^{-1} &= \eta_1 \,,
\end{split}
\end{equation}
which (assuming antiperiodic boundary conditions) is a translation to the right by one Majorana site accompanied by a sign flip. What does this imply for our $\mathbb{Z}_4$ symmetry? We combine $\mathcal{U}_{KW}^\prime$ with the reflection $\mathcal{R}$ into the symmetry operator $\mathcal{S}^\prime = \mathcal{R}\mathcal{U}_{KW}^\prime $. Its reflection action can be written compactly using $\widetilde{a} = 2N-a$ to denote the reflection symmetric of $a$ about the Majorana site $N$ (while, in the main text, the reflection is about site $N+1$). With this new convention, we find
\begin{equation}
    \begin{split}
        \mathcal{S}^\prime \,\eta_{2j-1}\, {\mathcal{S}^\prime}^{-1} &= i\mathcal{P}\,\eta_{\widetilde{2j-1}} \quad \text{for } a = 1,\dots,N,\\
        \mathcal{S}^\prime\, \eta_{2j}\, {\mathcal{S}^\prime}^{-1} &= - i\mathcal{P}\,\eta_{\widetilde{2j}} \quad \text{for } a = 1,\dots,N-1\,,\\
        \mathcal{S}^\prime \,\eta_{2N} \,{\mathcal{S}^\prime}^{-1} &= i\mathcal{P}\,\eta_{2N}\,.
    \end{split}
\end{equation}
Let us discuss open boundary conditions first. We observe that any local term involving the Majorana fermion $\eta_{2N}$ is mapped onto a non-local one and thus must be dropped from a local Hamiltonian. In this case, the symmetry-enhanced Hamiltonian must be defined with $h_{N} = 0$ and $h_1=h$ as compared to Eq.~\eqref{H_int}. Since it is now the Majorana fermion at site $N$ that is mapped onto itself, the interface should be constructed symmetrically around this site. To achieve this, we make the modifications $h_n = J_0$ and $h_{n+1}=J\,.$ The new set of coupling strengths is
\begin{align}
	&J_j = \begin{cases}
		J & j = 1, \,\dots,\, n-1 \\
		J_0 & j = n\\
		h & j = n+1,\,\dots,\,N-1
	\end{cases}\,, \quad
	h_j = \begin{cases}
		h & j = 1, \,\dots,\, n-1 \\
		J_0 & j = n \\
		J & j = n+1,\,\dots,\,N-1
	\end{cases}\,.
\end{align}
How are the zero modes affected by these changes? The trivial exact strong zero mode becomes $\eta_{2N}$. For any choice of the coupling constants, it partners with another exact strong zero mode composed of the Majorana operators with odd indices, which is localized at the edges if $J>h\,,$ or the interface if $h>J$ respectively. An additional interface-localized strong zero mode constructed from the Majorana operators with even indices, and another edges-bound strong zero mode with odd indices exist only in the regime where $J>h$, and their commutator with the Hamiltonian decreases exponentially with the system size. We thus recognize the same structure in our results as for our original definition of the Kramers-Wannier transformation with the regimes $J>h$ and $h>J$ being swapped. Finally, we comment on the closed chain: Here, we can use a mapping to relate the $\mathcal{S}$ symmetric fermion Hamiltonian to the $\mathcal{S}^\prime$ symmetric one. The transformation is implemented by a translation of every Majorana fermion by $N-1$ sites with antiperiodic boundary conditions, along with an interchange of the interface coupling parameters $J_0 \leftrightarrow J_0^\prime\,.$

We conclude that our findings do not show a qualitative dependence on the precise implementation of the Kramers-Wannier unitary.


\section{Bogoliubov-de Gennes equations for the dual-reflection interface}\label{app:BdG}
In Nambu space, a quadratic fermion Hamiltonian is represented by a matrix $\mathds{H}$, known as Bogoliubov-de Gennes (BdG) Hamiltonian,
\begin{align}
    \mathcal{H} = C^\dagger \mathds{H} C = \pqty{\hc{c}{1},\dots,\hc{c}{N},\op{c}{1},\dots,\op{c}{N}} \mathds{H} \pqty{\op{c}{1},\dots,\op{c}{N}, \hc{c}{1},\dots,\hc{c}{N}}^\top\,,
\end{align}
where $c_j$ and $c_j^\dagger$ are complex fermionic operators acting on site $j$ and obeying canonical fermionic anticommutation relations.
If $\mathds{U}$ is a unitary matrix that diagonalizes $\mathds{H}$, i.e., 
$
    \mathds{U}^\dagger \mathds{H} \mathds{U} $ $= \text{diag}(\dots)\, ,
    $
the Nambu spinor $\mathds{U}^\dagger C$ lists all 
fermionic eigenmodes, or quasiparticle excitations, of the problem. Formulated in terms of the matrix elements of $\mathds{U}\,,$ this diagonalization problem defines the so-called Bogoliubov-de Gennes equations. Once they are solved, we know the single-particle spectrum of the Hamiltonian and can use this information to study its many-body physics. In the following, we present the Bogoliubov-de Gennes equations for the open and closed $\mathcal{S}$ symmetric chain discussed in the main text, see Eqs.~\eqref{H_f} and \eqref{H_Majorana_closed}.

The BdG Hamiltonian of the transverse field Ising model (with generic non-uniform couplings) becomes block-tridiagonal after reshuffling the Nambu spinor as $\pqty*{\op{c}{1},\hc{c}{1},\dots,\op{c}{N},\hc{c}{N}}\,,$\vspace{-3pt}
\begin{align}
    \mathds{H}_{td} = A \mathds{H} A^{-1}\quad \text{with }
    A_{ij} = \begin{cases}
        \text{odd } i: & \delta_{i,\,j} \\
        \text{even } i: & \delta_{i/2+N,\,j}
    \end{cases}\quad .
\end{align}
The blocks are $2\cross2$-dimensional each, and the eigenvalue problem for this Hamiltonian amounts to finding two-component vectors $\bm{W}_{j,\mu} \in \mathds{C}^2 $ that solve the equations (written in terms of Pauli matrices $\tau^x,\,\tau^y$ and $\tau^z$)
\begin{align}\label{eq:TFIMscattering}
	-\frac{J_{j}}{2}\,(\tau^z + i \tau^y)\,\bm{W}_{j+1,\mu}  -\frac{J_{j-1}}{2}\,(\tau^z - i \tau^y)\,\bm{W}_{j-1,\mu} + h_j\, \tau^z\,\bm{W}_{j,\mu} = E_\mu \,\bm{W}_{j,\mu}\,,
\end{align}
for $j=1,\dots, \,N\,.$\footnote{The derivation of this equation is given as an exercise (Problem 3) in the review \cite{mbeng2024quantum}.} For an open chain, $\bm{W}_{N+1,\mu}$ and $J_N\,,$ as well as $\bm{W}_{0,\mu}$ and $J_0$ should be set to zero, while for a closed chain, we identify $\bm{W}_{N+1,\mu} \equiv \bm{W}_{1,\mu}$ and $J_N\equiv J_0\,.$ The index $\mu$ labels the eigenmodes and their energies $E_\mu\,,$ and $j$ refers to the site of a complex fermion (corresponding to a spin $1/2$) as before. 

The Majorana formulation of the BdG Hamiltonian $\mathds{H}_{td}$, which is more convenient for us, is obtained via $M\mathds{H}_{td}M^{-1}\,,$ with the block-diagonal matrix $M = \text{diag}\ \pqty{M_2,M_2,\dots,M_2}$, in which
\begin{align}
    M_2 = \mqty( 1 & 1\\ -i & i)\,,\quad M_2\, C_{j} = (\eta_{2j-1}, \eta_{2j})^\top\,.
\end{align}
If we apply this change of basis to the set of BdG equations \eqref{eq:TFIMscattering}, we find the new equations
\begin{align}
	J_{j}\, \mqty( 0 & 0\\ i & 0) \,\bm{V}_{j+1,\mu}  -J_{j-1}\, \mqty( 0 & i\\ 0 & 0)\,\bm{V}_{j-1,\mu} + h_j\, \mqty( 0 & i\\ -i & 0)\,\bm{V}_{j,\mu} = E_\mu \,\bm{V}_{j,\mu}\,.
\end{align}
In components, we write $\bm{V}_{j,\mu} = \pqty{V_{2j-1,\mu},\,V_{2j,\mu}}^\top$. $V_{2j-1,\mu}$ and $V_{2j,\mu}$ are the coefficients of the Majorana fermions $\eta_{2j-1}$ and $\eta_{2j}$ in the eigenmode $\Phi_\mu$ with energy $E_\mu\,,$
\begin{align}
    \Phi_\mu = \frac{1}{\sqrt{2}\,\mathcal{N}}\, (V_{1,\mu},\,V_{2,\mu},\,\dots,V_{2N,\mu}) \pqty{\eta_1,\,\eta_2,\,\dots,\eta_{2N}}^\top \,.
\end{align}
The resulting coupled linear equations are
\begin{equation}
\begin{split}
    -i J_{j-1}\, V_{2j-2,\mu} + i h_j\, V_{2j,\mu} &= E_\mu\, V_{2j-1,\mu}\, ,\\
    i J_{j}\, V_{2j+1,\mu} - i h_j\, V_{2j-1,\mu} &= E_\mu\, V_{2j,\mu}\,.
\end{split}
\end{equation}
Now, we will write these equations particularly for the Hamiltonian of the open (closed) dual-reflection symmetric chain by setting the coupling strengths to
\begin{align}
	&J_j = \begin{cases}
		J & j = 1, \,\dots,\, n-1 \\
		J_0 & j = n\\
		h & j = n+1,\,\dots,\,N-1\\
        0 \quad(-J_0^\prime) & j=N
	\end{cases}\,, \quad
	h_j = \begin{cases}
        0\quad (-J_0^\prime) & j=1 \\
		h & j = 2, \,\dots,\, n \\
		J_0 & j = n+1\\
		J & j = n+2,\,\dots,\,N
	\end{cases}\,.
\end{align}
We will also rename the indices $N+2,\,\dots\,,2N$ to reflection symmetric index pairs $a$ and $\widetilde{a}=2N+2-a\,,$ with $a = 1,\dots N\,,$ as in the main text. Thereby, in the bulk the equations are
\begin{align}\label{eq:Mcoeff_recursion_even}
        a=4,6,\dots,N: \qquad-i J\, V_{a-2,\mu} + i h\, V_{a,\mu} &= E_\mu\, V_{a-1,\mu}\,,\quad \nonumber\\
        -i h\, V_{\widetilde{a},\mu} + i J\, V_{\widetilde{a-2},\mu} &= E_\mu\, V_{\widetilde{a-1},\mu} \,, \vspace*{-20pt}
\end{align}
\begin{align}
        a=4,6,\dots,N-2: \qquad
        i J\, V_{a+1,\mu} - i h\, V_{a-1,\mu} &= E_\mu\, V_{a,\mu}\,,\qquad\nonumber\\
        i h\, V_{\widetilde{a-1},\mu} - i J\, V_{\widetilde{a+1},\mu} &= E_\mu\, V_{\widetilde{a},\mu}\, .
\end{align}
There are three equations involving the interface in the middle of the chain, $n=N/2\,,$
\begin{equation}
\begin{split}
    -i J_{0}\, V_{N,\mu} + i J_0\, V_{\widetilde{N},\mu} &= E_\mu\, V_{N+1,\mu}\qquad \text{for } j=n+1\,, \\
    i J_{0}\, V_{N+1,\mu} - i h\, V_{N-1,\mu} &= E_\mu\, V_{N,\mu}\qquad \text{for } j=n\,,\\
    i h\, V_{\widetilde{N-1},\mu} - i J_0\, V_{N+1,\mu} &= E_\mu\, V_{\widetilde{N},\mu}\qquad \text{for } j=n+1\,.
\end{split}
\end{equation}
At the left and right edge of the open chain, the following relations should hold:
\begin{align}
    iJ \, V_{3,\mu} = E_\mu\, V_{2,\mu}\,,\quad -iJ \, V_{\widetilde{3},\mu} &= E_\mu\, V_{\widetilde{2},\mu}\,,
\end{align}
and for the antiperiodic interface of the closed chain we require
\begin{equation}
\begin{split}
    i J_{0}^\prime\, V_{\widetilde{2},\mu} + i J_0\, V_{2,\mu} &= E_\mu\, V_{1,\mu}\qquad \text{for } j=N\,, \\
    i J\, V_{3,\mu} - i J_{0}^\prime\, V_{1,\mu} &= E_\mu\, V_{2,\mu}\qquad \text{for } j=1\,,\\
    -i J_{0}^\prime\, V_{1,\mu} - i J\, V_{\widetilde{3},\mu} &= E_\mu\, V_{\widetilde{2},\mu}\qquad \text{for } j=N\, .
\end{split}
\end{equation}
It is possible to express the BdG equations for the dual-reflection symmetric chain in terms of $V^\pm_{a,\mu} = V_{a,\mu} \pm (-1)^a V_{\widetilde{a},\mu}$ by taking the sum and difference of the equations involving the index $a$ and $\widetilde{a}$ respectively.\footnote{For example, Eq.~\eqref{eq:Mcoeff_recursion_even} is equivalent to
$ 
    -iJ \, V_{a-2,\mu}^\pm + ih\, V_{a,\mu}^\pm = E_\mu\ V_{a-1,\mu}^\pm
$.
} The equations for $V^+_{a,\mu}$ and $V^-_{a,\mu}$ decouple, which is a manifestation of the decoupling of $\mathcal{H}$ (and $\mathcal{H}^\circ$) into the $\xi^+$ and $\xi^-$ chains, see Sec.~\ref{sec:ferm_Z4}.

\section{Construction of strong zero modes from the BdG equations for the dual-reflection interface}\label{app:SZMs}
Every vector $\pqty*{ V_{1,\mu}, V_{2,\mu},\dots, V_{N,\mu},  V_{N+1,\mu}, V_{\widetilde{N},\mu},\dots, V_{\widetilde{2},\mu}} $ solving the BdG-equations specifies an operator that creates a particular eigenmode of the interface Hamiltonian with the energy $E_\mu\,,$ see Appendix~\ref{app:BdG}. These operators are linear combinations of Majorana fermions $\eta_a$. If there is an eigenmode with $E_\mu = 0$, the operator creating it commutes with the Hamiltonian. By construction, such a zero mode anticommutes with the fermion parity, which means that it is a Majorana strong zero mode (SZM), in this case even an exact SZM. Hence we can find strong zero modes by determining (approximate) zero-energy solutions to the BdG equations presented in Appendix~\ref{app:BdG}. 

We observe that at $E_\mu=0$ the BdG equations for the odd index and even index Majoranas decouple. Therefore we solve them separately. In the case $E_\mu=0\,,$ the interface equations dictate that (we will drop the index $\mu$ from now on)
\begin{align}
    J_0\,V_{N} &= J_0\, V_{\widetilde{N}}\,,\qquad V_{N-1}=J_0/h\, V_{N+1} = V_{\widetilde{N-1}}\,, \label{eq:Mcoeff_interface} \\
    J_0^\prime\,V_{2} &= -J_0^\prime\, V_{\widetilde{2}}\,,\hspace*{47pt} V_{3}=J_0^\prime/J\, V_{1} = - V_{\widetilde{3}}\,.\label{eq:Mcoeff_edge}
\end{align}
From the bulk equations, we get
\begin{align}
        a=4,6,\dots,N: & \qquad 
        V_{a-2} = \frac{h}{J}\, V_{a}\,,\quad V_{\widetilde{a-2}} = \frac{h}{J}\, V_{\widetilde{a}}\,,\label{eq:Mcoeff_recursion_zero_even}\\
        a=4,6,\dots,N-2: &  \qquad 
        V_{a-1} = \frac{J}{h}\, V_{a+1}\,,\qquad V_{\widetilde{a-1}} = \frac{J}{h}\, V_{\widetilde{a+1}}\,\,.\label{eq:Mcoeff_recursion_zero_odd}
\end{align}
These conditions clearly display reflection symmetry with respect to the site $N+1$. Note that the coefficient $V_1$, corresponding to the zero mode $\eta_1\, ,$ is not fixed by the BdG equations. The system of BdG equations for the (remaining) coefficients with odd indices does not have any nontrivial exact solution. Consequently, an exact strong zero mode must not contain Majorana fermions with any of the indices $3,5,\dots,2N-1\,.$ In some cases, it is nevertheless possible to construct approximate solutions involving those Majoranas, as discussed below.

\subsection{Strong zero modes of the open chain}
For the open chain, $J_0^\prime = 0\,,$ there is no constraint on the coefficients with even indices at the edge. Moreover, $\eta_1$ is an exact SZM on its own, corresponding to the solution $V_a=\delta_{a,1}\,,\, a=1,$ $\dots,2N\,.$
Locality and $\mathcal{S}$ symmetry together ensure the presence of this SZM, see Sec.~\ref{sec:robust_zero}.

\paragraph{Case $\bm{J>h}$:}
In the regime $J>h\,,$ recursion \eqref{eq:Mcoeff_recursion_zero_even} generates a series of coefficients that decays exponentially towards the edges of the chain.
Setting $V_N = V_{\widetilde{N}} = 1\,,$ we find an exact SZM which is composed of all the Majorana operators with even indices and localized around the interface. The nonzero coefficients are
\begin{align}
    V_{N-2j} = V_{\widetilde{N-2j}} = \pqty{\frac{h}{J}}^j\,,
\end{align}
for $j=0,\dots,N/2-1 \,.$
This solution corresponds to the following operator:
\begin{equation}
\begin{split}\label{eq:SZM_interface_even}
    \eta = { \frac{1}{\mathcal{N}}\sum_{j=0}^{n-1} \pqty{\frac{h}{J}}^{j}\,\frac{1}{\sqrt{2}}\pqty{\eta_{N-2j} + \eta_{\widetilde{N-2j}}} \overset{\text{Eq.~\eqref{eq:xi}}}{=} } \frac{1}{\mathcal{N}}\sum_{j=0}^{n-1} \pqty{\frac{h}{J}}^{j}\, \xi_{N-2j}^+\,,
\end{split}
\end{equation}
with the properties
\begin{align}
    \bqty{\mathcal{H},\, \eta}=0\,,\qquad \mathcal{S}\,\eta\,\mathcal{S}^{-1} = i\mathcal{P}\, \eta\,,
\end{align}
and with the normalization, chosen such that $\eta^2 = 1\,,$
\begin{equation}
    \mathcal{N}^2 = J^2\,\frac{ 1-\pqty{\frac{h}{J}}^N }{J^2 - h^2}\,.
\end{equation}
We observe that $\eta$ transforms under $\mathcal{S}$ in the same way as $\eta_1\,.$
Can we find an approximate solution using the remaining Majorana fermions with odd indices? According to recursion \eqref{eq:Mcoeff_recursion_zero_odd}, the largest coefficients have to be those at the edges, $V_{3} $ and $V_{\widetilde{3}}\,,$ whereby we would have to violate the condition $V_{3} = V_{\widetilde{3}} = 0$ at leading order to construct a non-zero operator. We cannot obtain another strong zero mode in this way.

\paragraph{Case $\bm{h>J}$:}
In the case $h>J\,,$ the series of even indices decays towards the interface and the largest coefficients are the ones at the edges; we choose $V_2 = V_{\widetilde{2}} = 1\,.$ Following recursion \eqref{eq:Mcoeff_recursion_zero_even} we end up with $V_N = V_{\widetilde{N}} = (J/h)^{N/2-1}\,,$ which satisfies condition \eqref{eq:Mcoeff_interface}. In total, the nonzero coefficients of our solution are
\begin{align}
    V_{2j} = V_{\widetilde{2j}} = \pqty{\frac{J}{h}}^{j-1}\,,
\end{align}
for $ j=1,\dots,N/2 \,.$
We conclude that there is an exact strong zero mode, symmetrically localized at the edges,
\begin{align}
    \eta =  \frac{1}{\mathcal{N}}\sum_{j=1}^{n} \pqty{\frac{J}{h}}^{j-1} \frac{1}{\sqrt{2}} \pqty{\eta_{2j} + \eta_{\widetilde{2j}}}  \overset{\text{Eq.~\eqref{eq:xi}}}{=}   \frac{1}{\mathcal{N}}\sum_{j=1}^{n} \pqty{\frac{J}{h}}^{j-1}\,\xi_{2j}^+\,,
\end{align}
which obeys
\begin{align}
     \bqty{\mathcal{H},\, \eta}=0\,, \qquad \mathcal{S}\,\eta\,\mathcal{S}^{-1} = i\mathcal{P}\, \eta\,,
\end{align}
and with the normalization
\begin{equation}
    \mathcal{N}^2 = h^2\,\frac{1-\pqty{\frac{J}{h}}^N }{h^2-J^2} \,.
\end{equation}
Note that the operator $\eta$ is the same for $J>h$ and $h>J$ when taking into account the different normalization factors, see also the footnote below Eq.~\eqref{eq:eta_edge_norm}.

Choosing $V_2 = - V_{\widetilde{2}} = 1$ as starting point instead leads us to a similar strong zero mode,
\begin{equation}
    \eta^{\prime\prime} =  \frac{1}{\mathcal{N}}\sum_{j=1}^{n} \pqty{\frac{J}{h}}^{j-1}\,\xi_{2j}^-\,,
\end{equation}
but it violates the interface condition due to the relative sign we introduced between $V_N$ and $V_{\widetilde{N}}\,.$ This sign also changes the transformation under $\mathcal{S}\,.$ Explicitly, $\eta^{\prime\prime}$ has the properties
\begin{equation}
    \bqty{\mathcal{H},\, \eta^{\prime\prime}} = -\frac{1}{\mathcal{N}} \pqty{\frac{J}{h}}^{n-1}\, 2\sqrt{2}iJ_0\, \xi_{N+1}^-\,, \qquad \mathcal{S}\,\eta^{\prime\prime}\,\mathcal{S}^{-1} = -i\mathcal{P}\, \eta^{\prime\prime}\,.
\end{equation}

Now, we turn to the odd indices.
As a consequence of $h>J$, the leading order condition for odd indices is the one at the interface, Eq.~\eqref{eq:Mcoeff_interface}. It is satisfied by the choice
\begin{align}
    V_{N+1}=1\,,\quad V_{N-1} = V_{\widetilde{N-1}} = \frac{J_0}{h}\,.
\end{align}
Next, using Eq.~\eqref{eq:Mcoeff_recursion_zero_odd}, we iteratively construct all remaining odd coefficients, ending up with $V_3 = V_{\widetilde{3}} = J_0/h \,\pqty{J/h}^{N/2-2}$. The remaining equation \eqref{eq:Mcoeff_edge} is now violated, but the mistake we make is suppressed exponentially with system size as required for a SZM. Explicitly, choosing the nonzero coefficients to be $V_{N+1}=1\,,$ and
\begin{equation}
   V_{N-(2j+1)} = V_{\widetilde{N-(2j+1)}} = \frac{J_0}{h}\pqty{\frac{J}{h}}^j 
\end{equation}
for $j=0,\dots,{N/2-2 }\,,$
using Eqs.~\eqref{eq:xi} and \eqref{eq:xi_special} we obtain the last SZM,
\begin{align}\label{eq:SZM_odd_open}
    \eta^\prime =  \frac{1}{\mathcal{N}^\prime}\bqty{\frac{1}{\sqrt{2}}\,\xi_{N+1}^- + \frac{J_0}{h} \sum_{j=0}^{n-2} \pqty{\frac{J}{h}}^{j}\,\xi_{N-(2j+1)}^{-}}\,,\qquad 
\end{align}
which is localized around the interface and fulfills
\begin{align}
     \bqty{\mathcal{H},\, \eta^\prime} = \frac{1}{\mathcal{N}^\prime}\frac{J_0}{h}\pqty{\frac{J}{h}}^{n-2}\, 2iJ\, \xi_2^- \,, \qquad \mathcal{S}\,\eta^\prime\,\mathcal{S}^{-1} = -i\mathcal{P}\, \eta^\prime\,,
\end{align}
with the normalization
\begin{equation}
    {\mathcal{N}^\prime}^2 = \frac{1}{2} + J_0^2\,\frac{1 - \pqty{\frac{J}{h}}^{N-2} }{h^2 - J^2} \,, 
\end{equation}

\subsection{Strong zero modes of the closed chain}
When $J_0^\prime\neq 0\,,$ the exact SZM $\eta_1$ is no longer present. In addition, also the system of equations for the coefficients with even indices no longer has a non-vanishing exact solution. Despite this, the closed chain hosts a pair of strong zero modes, as argued in the following. We inspect the recursion relations for the bulk to determine at which of the interfaces the coefficients should be largest, and use the corresponding constraints as a starting point for our iterative construction. Thereby, we can guarantee that the resulting operator commutes with the Hamiltonian to leading order, and does so to subleading order after taking into account further bulk equations. The condition at the opposite interface is then violated, but with the exponential-in-size scaling behaviour.

\paragraph{Case $\bm{J>h}$:}
From Eq.~\eqref{eq:Mcoeff_recursion_zero_even}, we infer that an approximate solution for even indices should satisfy condition \eqref{eq:Mcoeff_interface} of the periodic interface, around which it will be localized.
The corresponding SZM is identical to the interface mode of the open chain for $J>h$, Eq.~\eqref{eq:SZM_interface_even}, now with the properties
\begin{align}
    \bqty{\mathcal{H}^\circ,\, \eta}= \frac{1}{\mathcal{N}}\pqty{\frac{h}{J}}^{n-1}\,2\sqrt{2}iJ_0^\prime\, \xi_1^+\,,\qquad \mathcal{S}\,\eta\,\mathcal{S}^{-1} = i\mathcal{P}\, \eta\,.
\end{align}
An approximate solution using odd indices must satisfy Eq.~\eqref{eq:Mcoeff_edge} at the antiperiodic interface, as follows from Eq.~\eqref{eq:Mcoeff_recursion_zero_odd}. Therefore, we make the initial choice $V_3 = - V_{\widetilde{3}} = J_0^\prime/J\,,$ which leads to the nonzero coefficients
\begin{align}
    V_1 = 1\,,\quad V_{2j-1} = -V_{\widetilde{2j-1}} = \frac{J_0^\prime}{J} \pqty{\frac{h}{J}}^{j-2} 
\end{align}
for $j=2,\dots,N/2 \,.$
The operator specified by these coefficients is localized around the antiperiodic interface. Using Eqs.~\eqref{eq:xi} and \eqref{eq:xi_special}, it reads
\begin{align}
    \eta^\prime =  \frac{1}{\mathcal{N}^\prime}\bqty{ \frac{1}{\sqrt{2}}\, \xi_1^+ + \frac{J_0^\prime}{J}\sum_{j=2}^{n} \pqty{\frac{h}{J}}^{j-2}\,\xi_{2j-1}^{+}}\,,
\end{align}
and it obeys
\begin{align}
    \bqty{\mathcal{H}^\circ,\, \eta^\prime}= \frac{1}{\mathcal{N}^\prime}\frac{J_0^\prime}{J}\pqty{\frac{h}{J}}^{n-2}\pqty{-2ih\,\xi_{N}^+}\,\,,\qquad \mathcal{S}\,\eta^\prime\,\mathcal{S}^{-1} = i\mathcal{P}\, \eta^\prime\,,
\end{align}
with the normalization
\begin{equation}
     {\mathcal{N}^\prime}^2 = \frac{1}{2} + J_0^\prime\,\frac{{ 1 - \pqty{\frac{h}{J}}^{N-2} }}{J^2 - h^2}\,.
\end{equation}
Both interface SZMs transform alike under $\mathcal{S}\,.$

\paragraph{Case $\bm{h>J}$:}
We can map the above Hamiltonian $\mathcal{H}^\circ \equiv \mathcal{H}^\circ(J,h,J_0,J_0^\prime)$ onto a Hamiltonian with exchanged coupling strengths $\mathcal{H}^\circ(J\leftrightarrow h,\,J_0\leftrightarrow J_0^\prime)$ by the transformation
\begin{align}
    \eta_{a} \rightarrow \begin{cases}
    \eta_{N+a} & a\leq N\\
    -\eta_{a-N} & a > N
    \end{cases}\qquad \text{for } a=1,\dots,\,2N\,,
\end{align}
which (up to a sign) corresponds to a translation of the Majorana operators by $N$ sites modulo $2N\,.$ It acts on $\xi_a^\pm$ as
\begin{align}
    \xi_a^p \rightarrow p\,(-1)^{a+1}\,\xi_{N+2-a}^{-p} \qquad \text{with } p=\pm \,.
\end{align}
The SZMs of $\mathcal{H}^\circ(J\leftrightarrow h,\,J_0\leftrightarrow J_0^\prime)$ follow directly from those derived in the subsection discussing $J>h$ using this mapping and exchanging $J\leftrightarrow h,\,J_0\leftrightarrow J_0^\prime$ in the given expressions for the SZMs. For completeness, we state the explicit formulas. The operator localized around the antiperiodic interface is
\begin{align}
    \eta  =  \frac{1}{\mathcal{N}}\sum_{j=1}^{n} \pqty{\frac{J}{h}}^{j-1}\,\xi_{2j}^- \,,
\end{align}
with the properties
\begin{align}
     \bqty{\mathcal{H}^\circ,\, \eta}= \frac{1}{\mathcal{N}}\pqty{\frac{J}{h}}^{n-1}\,\pqty{-2\sqrt{2}iJ_0\, \xi_{N+1}^-}\,, \qquad \mathcal{S}\,\eta\,\mathcal{S}^{-1} = - i\mathcal{P}\, \eta\,,
\end{align}
and the normalization
\begin{equation}
   \mathcal{N}^2 = h^2\,\frac{ 1-\pqty{\frac{J}{h}}^N }{h^2 - J^2}\,.
\end{equation}
The operator of the SZM at the periodic interface is identical with the interface mode of the open chain for $h>J$, Eq.~\eqref{eq:SZM_odd_open}, now obeying
\begin{align}
    \bqty{\mathcal{H}^\circ,\, \eta^\prime}= \frac{1}{\mathcal{N}^\prime}\frac{J_0}{h}\pqty{\frac{J}{h}}^{n-2}\,2iJ\xi_2^-\,,\qquad \mathcal{S}\,\eta^\prime\,\mathcal{S}^{-1} = - i\mathcal{P}\, \eta^\prime\,.
\end{align}

\section{Exact Floquet Majorana zero mode of the dual-reflection interface}\label{app:FMZM}

In this section we derive the explicit form of the exact Floquet Majorana zero mode \eqref{eq:interface FMZM} presented in Sec.~\ref{sec:quantum circuit}. To this end, following the method of Appendix III in \cite{Google2022}, we first construct the exact Floquet mode for a Majorana chain where the first Majorana fermion decouples from the Floquet unitary. Next, using this result and the representation illustrated in Fig.~\ref{fig:double_chain}, we derive the exact Floquet zero mode of the dual-reflection interface model investigated in this paper.

\subsection{Floquet zero modes of a Majorana chain}
In this section, we study the evolution of a Majorana fermion chain of length $2n$ under a unitary Floquet operator $\mathcal{U}_F$ of the form
\begin{equation}\label{eq:Floquet_unitary}
	\mathcal{U}_F = \prod_{j=1}^{n} \exp{-\frac{\varphi_j}{2}\, \eta_{2j-1}\eta_{2j}}\, \prod_{j=1}^{n-1} \exp{-\frac{\theta_j}{2}\, \eta_{2j}\eta_{2j+1}} \,,
\end{equation}
where $\eta_a$ is a Majorana operator as in the main text (Eq.~\eqref{eq:eta_def}).
Its inverse is
\begin{equation}\label{eq:Floquet unitary -1}
	\mathcal{U}_F^{-1} = \mathcal{U}_F^\dagger = \prod_{j=1}^{n-1} \exp{\frac{\theta_j}{2}\, \eta_{2j}\eta_{2j+1}}\,\prod_{j=1}^{n} \exp{\frac{\varphi_j}{2}\, \eta_{2j-1}\eta_{2j}}\,,
\end{equation}
As defined in Sec.~\ref{sec:quantum circuit}, we call a Floquet strong zero mode a Majorana operator $\Psi$, 
\begin{equation}\label{eq:Psi}
	\Psi = \sum_{j=1}^{n} \psi_{2j-1}\,\eta_{2j-1} + \psi_{2j}\, \eta_{2j} \,,
\end{equation}
which is odd under fermion parity $\mathcal{P}$ and invariant under Floquet evolution, i.e. commutes with the evolution operator $\mathcal{U}_F$,
\begin{equation} \label{eq:FZMcondition}
	\mathcal{U}_F^{-1}\,\Psi\,\mathcal{U}_F = \Psi\,.
\end{equation}
An explicit calculation of the right hand side of this equality using the ansatz \eqref{eq:Psi} yields
\begin{equation*}\label{eq:Psi evolution}
	\begin{split}
		&\mathcal{U}_F^{-1}\,\Psi\,\mathcal{U}_F \\
		&= \big( \cos(\varphi_1)\, \psi_1 + \sin(\varphi_1)\, \psi_2\big)\,\eta_1 + \big(-\sin(\varphi_n)\,\psi_{2n-1} + \cos(\varphi_n)\,\,\psi_{2n}\big)\,\eta_{2n}\\
		& + \sum_{j=1}^{n-1} \big( -\sin(\varphi_{j})\cos(\theta_{j})\,\psi_{2j-1} + \cos(\varphi_{j})\cos(\theta_{j})\,\psi_{2j} \\[-10pt]
		& \hspace*{36pt}+ \cos(\varphi_{j+1})\sin(\theta_j)\,\psi_{2j+1} + \sin(\varphi_{j+1})\sin(\theta_{j})\,\psi_{2j+2}  \big)\, \eta_{2j} \\
		&\hspace*{19pt}+ \big( \sin(\varphi_{j})\sin(\theta_{j})\,\psi_{2j-1} - \cos(\varphi_j)\sin(\theta_{j})\,\psi_{2j} \\
		& \hspace*{36pt} + \cos(\varphi_{j+1})\cos(\theta_j)\,\psi_{2j+1} + \sin(\varphi_{j+1})\cos(\theta_j)\,\psi_{2j+2}  \big)\, \eta_{2j+1}
	\end{split}
\end{equation*}
If we set this equal to Eq.~\eqref{eq:Psi} and compare the coefficients in front of each Majorana operator $\eta_a$, we obtain the boundary conditions
\begin{align}
	\cos(\varphi_1)\, \psi_1 + \sin(\varphi_1)\, \psi_2 &= \psi_1\,,\label{eq:bdy1_general}\\
	-\sin(\varphi_n)\,\psi_{2n-1} + \cos(\varphi_n)\,\,\psi_{2n} &= \psi_{2n}\,,\label{eq:bdy2N_general}
\end{align}
as well as $2(n-1)$ further constraints on $\psi_{2j-1}$ and $\psi_{2j}$, which we can write in the following way:
\begin{equation} \label{eq:bulk_condition_general}
	\begin{split}
		&\mqty(1-\cos(\varphi_{j+1})\, \cos(\theta_j) & -\sin(\varphi_{j+1})\,\cos(\theta_j)\\ \cos(\varphi_{j+1})\,\sin(\theta_j) & \sin(\varphi_{j+1})\, \sin(\theta_j))\, \mqty(\psi_{2j+1}\\ \psi_{2j+2}) \\
		&=\mqty(\sin(\varphi_j)\sin(\theta_{j}) & -\cos(\varphi_j)\,\sin(\theta_{j})\\ \sin(\varphi_j)\,\cos(\theta_{j}) & 1-\cos(\varphi_j)\,\cos(\theta_{j}))\, \mqty(\psi_{2j-1}\\\psi_{2j})\quad \text{for } j=1\,,\dots\,, n-1\,.
	\end{split}
\end{equation}

Motivated by the form of the reflection-dual open chain illustrated in Fig. \ref{fig:double_chain}, as will become clear in the second part of this appendix, we now consider the Majorana chain in the case in which the first Majorana fermion decouples from the unitary evolution. Specifically,
we choose the phases
\begin{equation}
	\begin{split}
		\varphi_j &= \begin{cases}
			0 & j=1\\
			\varphi & j=2,\,\dots\,,n
		\end{cases}\\
		\theta_j &= \theta \quad \forall \,j\,.
	\end{split}
\end{equation}
In the bulk where $\varphi_j = \varphi$ and $\theta_j = \theta\,,$ the previous equation can be written as
\begin{equation}\label{eq:T-1eq}
	\mqty(\psi_{2j-1}\\\psi_{2j}) = T^{-1}\mqty(\psi_{2j+1}\\ \psi_{2j+2}) \qquad \text{for } j=2,\dots,n-1\,,
\end{equation}
where we have introduced the transfer matrix
\begin{equation}\label{eq:T}
	T^{-1} =  \frac{1}{\sin(\theta)} \mqty( \frac{1}{\sin(\varphi)}\pqty{1+ \cos(\varphi)^2 -2\,\cos(\theta)\cos(\varphi)} &\cos(\varphi) - \cos(\theta)\\ \cos(\varphi)-\cos(\theta) & \sin(\varphi) ) \,.
\end{equation}
Assuming that $\varphi_1=0$, the boundary condition \eqref{eq:bdy1_general} at the left end is satisfied trivially. The conditions stated in Eq.~\eqref{eq:bulk_condition_general} become
\begin{equation}\label{eq:bdy-1}
	\mqty(1-\cos(\varphi)\cos(\theta) & -\sin(\varphi) \cos(\theta)\\ \cos(\varphi) \sin(\theta) & \sin(\varphi) \sin(\theta))\, \mqty(\psi_{3}\\ \psi_{4}) = \mqty( 0 & -\sin(\theta)\\ 0 & 1-\cos(\theta))\, \mqty(\psi_{1}\\\psi_{2})\,,
\end{equation}
and need to be satisfied together with Eqs.~\eqref{eq:bdy2N_general} and \eqref{eq:T-1eq} by a Floquet zero mode. Now, we construct such an operator.

The transfer matrix equations \eqref{eq:T-1eq} can be solved by diagonalizing $T$ as follows: If $\Psi_0$ is an eigenvector of the $T$-matrix, 
\begin{equation}
	T^{-1}\,\Psi_0 = \lambda_0 \Psi_0\,,
\end{equation}
and if we set
\begin{equation}\label{eq:power expansion generic}
	\mqty(\psi_{2j-1}\\\psi_{2j}) = {\lambda_0}^{n-j}\,\Psi_0 \qquad \text{for } j=1,\dots,n\,,
\end{equation}
we find
\begin{equation}
	T^{-1}\, \mqty(\psi_{2j+1}\\\psi_{2j+2}) = {\lambda_0}^{n-j-1} \, T^{-1}\,\Psi_0 = {\lambda_0}^{n-j} \Psi_0 = \mqty(\psi_{2j-1}\\\psi_{2j})\,.
\end{equation}
The eigensystem of $T^{-1}$ is
\begin{equation}\label{eq:eigsys-1}
	\Bqty{\frac{1}{\lambda_0},\,\mqty(\cos(\varphi/2)\\\sin(\varphi/2))}\,,\quad\Bqty{\lambda_0,\,\mqty(-\sin(\varphi/2)\\\cos(\varphi/2))}\,,\qquad \text{with } \lambda_0 = \frac{\tan(\varphi/2)}{\tan(\theta/2)}\,.
\end{equation}
In the regime $\lambda_0 < 1$, let us make the following ansatz for the Floquet SZM \eqref{eq:Psi}:
\begin{equation}\label{eq:psi right}
	\mqty(\psi_{2n-1}\\\psi_{2n}) = \mqty(-\sin(\varphi/2)\\\cos(\varphi/2))\,.
\end{equation}
The recursion \eqref{eq:T-1eq} then yields coefficients that decay from the right end to the left end as
\begin{equation}\label{eq:power expansion -1}
	\mqty(\psi_{2j-1}\\\psi_{2j}) = {\lambda_0}^{n-j}\mqty(-\sin(\varphi/2)\\\cos(\varphi/2)) \qquad \text{for } j=1,\dots,n-1\,.
\end{equation}
We would like this ansatz to satisfy the boundary condition at the right edge, eq.~\eqref{eq:bdy2N_general}, for which we correctly find the identity 
\begin{equation}\label{eq:trigonometric}
	\cos(\varphi) \cos(\varphi/2) + \sin(\varphi) \sin(\varphi/2) = \cos(\varphi/2)\,.
\end{equation}
For generic $\varphi_1$, our ansatz violates the boundary condition \eqref{eq:bdy1_general} at order ${\lambda_0}^{n-1}$ and becomes a Floquet SZM in the limit $n\to\infty$. In the following, we show that this ansatz even yields an exact Floquet SZM if we set $\varphi_1$ to zero. If we insert the coefficients \eqref{eq:power expansion -1} into Eq.~\eqref{eq:bdy-1}, we obtain
\begin{equation}
	\mqty(1-\cos(\varphi)\cos(\theta) & -\sin(\varphi) \cos(\theta)\\ \cos(\varphi) \sin(\theta) & \sin(\varphi) \sin(\theta))\, \mqty(\psi_{2n-1}\\ \psi_{2n}) = \lambda_0\, \psi_{2n} \mqty(-\sin(\theta)\\ 1-\cos(\theta))\,,
\end{equation}
where $\lambda_0$ is given in eq.~\eqref{eq:eigsys-1} and our ansatz for $\psi_{2n-1}, \psi_{2n}$ in eq.~\eqref{eq:psi right}. Both equalities (the upper and the lower component of the vector identity) are satisfied if we make these replacements. Thereby, the following Majorana mode commutes with the Floquet evolution operator exactly:
\begin{align}\label{eq:FMZM}
	\Psi = \frac{1}{\mathcal{N}}\,\bqty{ {\lambda_0}^{n-1} \cos(\frac{\varphi}{2}) \, \eta_{2} + \sum_{j=2}^{n} {\lambda_0}^{n-j} \Big(-\sin(\frac{\varphi}{2})\, \eta_{2j-1} + \cos(\frac{\varphi}{2})\, \eta_{2j} \Big)  }\,,
\end{align}
with the normalization
\begin{align}
	\mathcal{N}^2 &= 2 \cos(\frac{\theta}{2})^2 \frac{1-{\lambda_0}^2}{1 + \cos(\theta) - 2\cos(\varphi/2)^2 \,{\lambda_0}^{2n}}\,.
\end{align}
It is localized near the right (left) boundary if $\lambda_0 < 1$  ($\lambda_0 > 1$). Clearly, $\Psi$ comes together with the Floquet Majorana strong zero mode
\begin{equation}
	\Psi_1 = \eta_1\,,
\end{equation}
which does not appear in the evolution operator.

\subsection{Floquet zero mode of the dual-reflection interface}
In the following, we derive an exact Floquet strong zero mode of the Floquet circuit with dual-reflection symmetry in the Majorana representation. 
First, we write the circuit operator $U= U_h \, U_{J_0} \, U_J$, see Eq.~\eqref{eq:U_dig}, in terms of Majorana operators. As in the main text, we set $\varphi = \pi h \,$, $\theta = \pi J\,$, and $\alpha = \pi J_0\,$.
To this end, it is convenient to rewrite $U$ combining exponents, e.g. for $U_h\,$:
\begin{equation}
\begin{split}
	U_h&= \prod_{j=n+1}^{N-1} \exp\left(-i\frac{\pi h}{2} X_j X_{j+1}\right) \prod_{j=2}^{n} \exp\left(-i\frac{\pi h}{2} Z_j\right) \\
	&= \exp\bqty{-i\frac{\pi h}{2}\pqty{\sum_{j=n+1}^{N-1}X_j X_{j+1} + \sum_{j=2}^n Z_j}}\,.
\end{split}
\end{equation}
After Jordan-Wigner transformation into Majorana fermions $\xi^+$ and $\xi^-$, this becomes
\begin{equation}
	\mathcal{U}_h = \exp\bqty{-\frac{\pi h}{2}\sum_{j=2}^{n}\pqty{\xi^+_{2j-1}\xi^+_{2j} + \xi^-_{2j-1}\xi^-_{2j}}}\,.
\end{equation}
Since all summands in the exponent commute with each other, we arrive at the following expressions for $\mathcal{U}_h$ and analogously $\mathcal{U}_J$ and $\mathcal{U}_{J_0}\,$:
\begin{align}
	\mathcal{U}_h&= \prod_{j=2}^{n} \exp{-\frac{\pi h}{2}\, \xi^+_{2j-1}\xi^+_{2j}}\, \prod_{j=2}^{n} \exp{-\frac{\pi h}{2}\, \xi^-_{2j-1}\xi^-_{2j}}\,,\\
	\mathcal{U}_{J}&= \prod_{j=1}^{n-1} \exp{-\frac{\pi J}{2}\, \xi^+_{2j}\xi^+_{2j+1}}\, \prod_{j=1}^{n-1} \exp{-\frac{\pi J}{2}\, \xi^-_{2j}\xi^-_{2j+1}}\,,\\
	\mathcal{U}_{J_0}&= \exp\Bqty{-\frac{\pi J_0}{\sqrt{2}}\,\xi^-_{N}\xi^-_{N+1}}\,.
\end{align}
The Majorana representation of the operator $U$ is
\begin{align}
	\mathcal{U} = \mathcal{U}_h\,\mathcal{U}_{J_0}\,\mathcal{U}_J \,.
\end{align}
Using that Majoranas belonging to the $\xi^+$ chain anticommute with those on the $\xi^-$ chain, we can even write the evolution operator as a product of the mutually commuting operators
\begin{equation}
	\mathcal{U} = \mathcal{U}_+\,\mathcal{U}_-\,,\qquad \bqty{\mathcal{U}_+,\,\mathcal{U}_-}=0\,,
\end{equation}
in which
\begin{align}
    \mathcal{U}_{+} &=  \prod_{j=2}^{n} \exp{-\frac{\pi h}{2}\, \xi^+_{2j-1}\xi^+_{2j}} \, \prod_{j=1}^{n-1} \exp{-\frac{\pi J}{2}\, \xi^+_{2j}\xi^+_{2j+1}} \,,\label{eq:U+}\\
	\mathcal{U}_- &= \prod_{j=2}^{n} \exp{-\frac{\pi h}{2}\, \xi^-_{2j-1}\xi^-_{2j}}\, \exp\Bqty{-\frac{\pi J_0}{\sqrt{2}}\,\xi^-_{N}\xi^-_{N+1}} \, \prod_{j=1}^{n-1} \exp{-\frac{\pi J}{2}\, \xi^-_{2j}\xi^-_{2j+1}} \,.
\end{align}
With these definitions, the above Floquet SZM $\Psi\,$, see Eq.~\eqref{eq:FMZM}, exactly commutes with $\mathcal{U}_+$ if we replace $\eta_a$ by $\xi^+_a\,$, and assume that $\varphi = \pi h$ and $\theta = \pi J\,$.
Consider now a Trotter evolution of $\Psi$ with $\mathcal{U}\,.$ Since $\Psi$ belongs to the $\xi^+$ chain, even $\mathcal{U}_-$ commutes with it, whereby we find that this Majorana mode is an exact solution of
\begin{equation}
	\mathcal{U}^{-1}\,\Psi\,\mathcal{U} = \Psi\,.
\end{equation}

\end{appendix}





\bibliography{lib}


\end{document}